\documentclass{emulateapj}
\usepackage{graphicx}    
\usepackage{subfigure}
\usepackage{color}
\newcommand{\Msun}{h^{-1} M_{\odot}}

\submitted{}

\begin{document} 
\title{Merger Histories of Galaxy Halos and Implications for Disk Survival}
\author{Kyle R. Stewart\altaffilmark{1}, James S. Bullock\altaffilmark{1}, 
Risa H. Wechsler\altaffilmark{2}, Ariyeh H. Maller\altaffilmark{3}, and Andrew R. Zentner\altaffilmark{4}}

\altaffiltext{1}{Center for Cosmology, Department of Physics and Astronomy, The University of California at Irvine, Irvine, CA, 92697, USA}
\altaffiltext{2}{Kavli Institute for Particle Astrophysics \& Cosmology, Physics Department, and Stanford Linear Accelerator Center, Stanford University, Stanford, CA 94305, USA}
\altaffiltext{3}{Department of Physics, New York City College of Technology, 300 Jay St., Brooklyn, NY 11201, USA}
\altaffiltext{4}{Department of Physics and Astronomy, University of Pittsburgh, Pittsburgh, PA 15260, USA}

\begin{abstract} {We study the merger histories of galaxy dark matter
    halos using a high resolution $\Lambda$CDM $N$-body simulation.
    Our merger trees follow $\sim 17,000$ halos with masses $M_0 =
    (10^{11} - 10^{13}) \Msun$ at $z = 0$ and track accretion events
    involving objects as small as $m \simeq 10^{10} \Msun$.  We find
    that mass assembly is remarkably self-similar in $m/M_0$, and
    dominated by mergers that are $\sim 10 \%$ of the final halo mass.
    While very large mergers, $m \gtrsim 0.4 \, M_0$, are quite rare,
    sizeable accretion events, $m \sim 0.1 \, M_0$, are common.  Over
    the last $\sim 10$ Gyr, an overwhelming majority ($\sim 95\%$) of
    Milky Way-sized halos with $M_0 = 10^{12} \Msun$ have accreted at
    least one object with greater total mass than the Milky Way disk
    ($m > 5 \times 10^{10} \Msun$), and approximately $70 \%$ have
    accreted an object with more than twice that mass ($m > 10^{11}
    \Msun$). Our results raise serious concerns about the survival of thin-disk
    dominated galaxies within the current paradigm for galaxy
    formation in a $\Lambda$CDM universe.  In order to achieve a $\sim 70 \%$
    disk-dominated fraction in Milky Way-sized $\Lambda$CDM halos, mergers
    involving $m \simeq 2 \times 10^{11} \Msun$ objects must not
    destroy disks.  Considering 
    that most thick disks and bulges contain old stellar populations, 
    the situation is even more restrictive:
    these mergers must not heat disks or drive gas into their centers to create young bulges.}

\end{abstract}

\section{Introduction}
\label{Introduction}

In the cold dark matter (CDM) model of structure formation, dark
matter halos form via the continuous accretion of smaller systems
\citep{Peebles82,Blumenthal84,Davis85,FakhouriMa07,NeisteinDekel08,Cole08}.  
Mergers of the type predicted
can help explain many properties of the observed universe.  Major
mergers are believed to play an important role in shaping the Hubble
sequence
\cite[][]{tt72,barnes88,hernquist93,nb03,kb05,cox06a,robertson06b,robertson06a,r06,maller06,jesseit07,bournaud07}
and triggering star formation and AGN activity
\cite[][]{mihos96,kolatt99,cox06,woods06,barton07}.  Minor mergers may
help explain the origin of thick disks
\citep{quinn93,walker_etal96,abadi_03,brook04,dalcanton_etal05,k07,HayashiChiba06},
cause anti-truncation \citep{younger07}, and produce extended diffuse
light components around galaxies
\citep{johnston96,helmi99,bkw01,bj05,purcell07,bell07}.  However,
there is lingering concern that mergers are too common in CDM
cosmologies for thin disk-dominated systems to survive
\citep[][]{toth_ostriker92,wyse01,kormendy05,kautsch06}.  In this
paper we present the merger statistics necessary for addressing this
issue.

The formation of disk galaxies within hydrodynamic simulations in
hierarchical CDM cosmologies has proven problematic
\citep[e.g.][]{ns00}.  While there have been some successes in forming
galaxies with disks in cosmological simulations
\citep[][]{abadi_03,sommer_larsen_etal03,brook04,robertson04,kaufmann07a,governato_etal07},
the general problem is far from resolved. The resultant disks are
often fairly thick and accompanied by large bulges, and the systems
that form disks tend to have special merger histories.  The resultant 
thick disk and bulge stars also tend towards a broad range of stellar ages, instead
of being dominated by predominantly old stars.  
Moreover, the
successes depend strongly on effective models that describe physics on
scales far below the simulation resolution, which are poorly
understood.
 Given the  current
difficulties in  understanding {\it ab  initio} disk formation, one 
can consider a  less ambitious, but more well-posed question.   
Even if disk galaxies
can form within   CDM halos, can they survive the predicted merger 
histories?

Unfortunately, the prevalence of thin-disk or even disk-dominated
galaxies in the universe is difficult to quantify with large
observational samples.  Some promising approaches use asymmetry vs.
concentration to define morphological type
\citep[e.g.,][]{Ilbert06alt}, and some use a combination color and
concentration indicators \cite[e.g.][]{Choi07, Park07}.  Despite the
wide range of definitions, the general consensus in the literature is
that $\sim 70 \%$ of $\sim 10^{12} \Msun$ halos host disk-dominated,
late-type galaxies
\citep[e.g.][]{Weinmann06,vandenBosch06,Ilbert06alt,Choi07,Park07}.
We adopt this number for the sake of discussion in what follows, but
none of our primary results on merger statistics depend on this
number.

Also relevant to the discussion of galaxy merger histories is the
prevalence of pure disk galaxies in the universe.  \cite{kautsch06}
have compiled a statistically meaningful sample of edge-on disk
galaxies and found that $\sim 16 \%$ of these objects are ``simple
disks'' with no observable bulge component.  In principle, this
statistic places strong constraints on the merger histories of
galaxies.  Moreover, a large fraction of disk galaxies with bulges
contain pseudo-bulges, which may be the products of secular processes
and not the remnants of an early merger event
\citep[e.g.][]{kk04,carollo07}.  These cases provide further
motivation to quantify the predicted merger histories of galaxy halos
in the favored CDM cosmology.

Here we use a large dissipationless cosmological $\Lambda$CDM N-body
simulation to track the merger histories of an ensemble of $\sim
17,000$ halos with $z=0$ masses $M_0 = 10^{11} - 10^{13} \Msun$.  We
focus on halos of fixed mass at $z=0$, and concentrate specifically on
Milky Way-sized systems, $M_0 \simeq 10^{12} \Msun$.  We categorize
the accretion of objects as small as $m \simeq 10^{10} \Msun$ and
focus on the infall statistics into {\em main progenitors} of $z=0$
halos as a function of lookback time.  As discussed below, the {\em
  main progenitor} is defined to be the most massive progenitor of a
$z=0$ halo tracked continuously back in time.

A {\em merger} is defined here to occur when an infalling halo first
crosses within the virial radius of the main progenitor.  In most
cases we do not track subhalo evolution after accretion.  We have
chosen to track mergers in this way in order to provide a robust
prediction.  An understanding of an accreted halo's subsequent orbital
evolution and impact with the central disk region is essential for any
complete understanding of {\em galaxy} merger statistics.  However,
this evolution will be sensitive to the baryonic distribution within
both the main progenitor halo and the satellites themselves.  The {\em
  halo} merger rate we present is a relatively clean measure that can
be used as a starting point for more detailed investigations of
galaxy--galaxy encounters.  Still, it is worth pointing out that for
most of the mergers we consider, impacts with the central disk region
should occur relatively shortly after accretion.  As we show in the
Appendix, events with $m \gtrsim 0.1 M_0$ typically happen at a
redshift $z$ when the main progenitor mass, $M_z$, is significantly
smaller than $M_0$, such that the merger ratio is fairly large $m/M_z
\gtrsim 0.2$.  Therefore, even ignoring the enhanced orbital decay
that will be caused by a central disk potential, the dynamical
friction decay times are expected to be short for these events, with
central impacts occurring within $\tau \lesssim 3$ Gyr for typical
orbital parameters \citep{bk07,z05,zb03}.  As discussed in conjunction
with Figure 5 in \S 3, destruction times of $\sim 3$ Gyr are
consistent with our measurements of subhalo evolution.

The outline of this paper is as follows.  In \S \ref{Simulation} we
discuss the numerical simulations used and the method of merger tree
construction.  In \S \ref{Results} we present our principle results,
which characterize the accretion mass functions of halos and the
fraction of halos with mergers as a function of lookback time.  In \S
\ref{Discussion} we discuss these results in reference to the problem
of disk survival in a hierarchical universe, and we summarize our main
conclusions in \S \ref{Conclusion}.

\begin{figure*}[t!]
  \begin{center}
  \subfigure{\includegraphics[width=.45\textwidth]{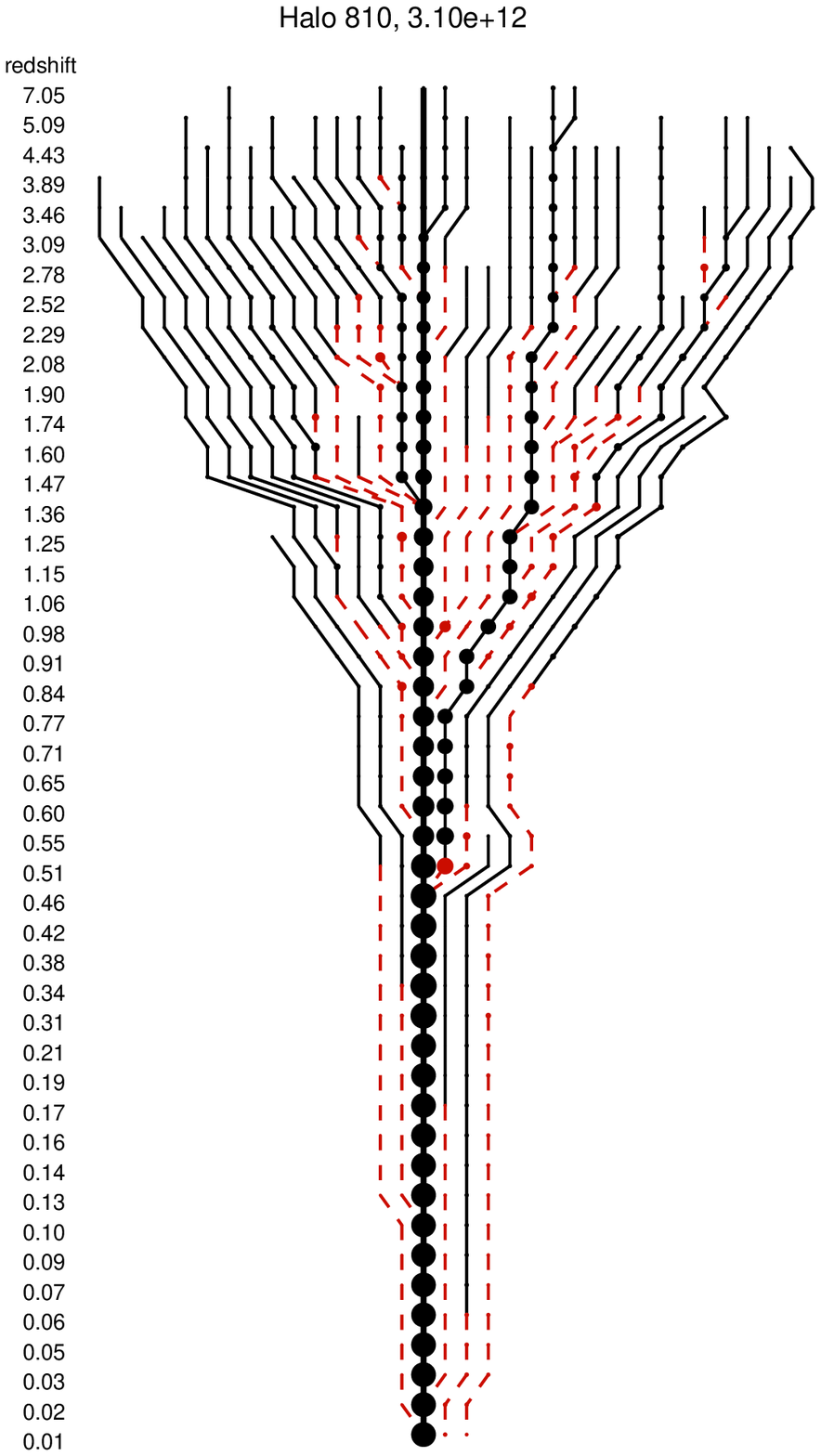}} 
  \subfigure{\includegraphics[width=.45\textwidth]{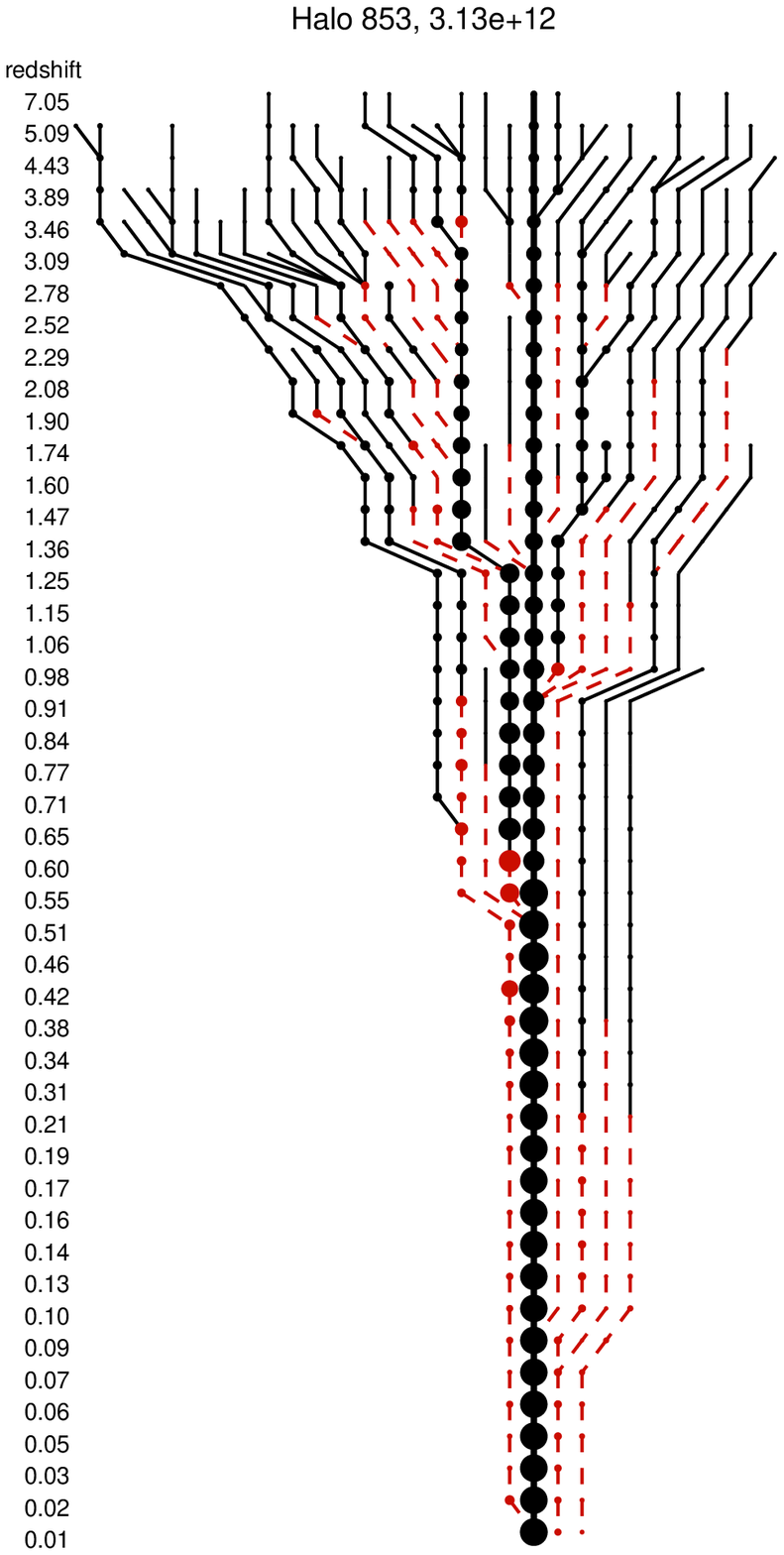}} 
  \caption{Sample merger trees, for halos with $M_{0} \simeq 10^{12.5}
  \Msun$.   Time progresses downward,  with the redshift $z$
  printed on the left hand side.  
 The  bold, vertical line at the center corresponds
 to   the main progenitor,  with  filled  circles proportional to the
  radius of  each halo.  The minimum  mass halo shown in  this diagram
  has   $m =   10^{9.9}  \Msun$.
 Solid (black) and    dashed (red) lines and    circles
  correspond   to  isolated field  halos,  or  subhalos, respectively.
The dashed (red) lines that do not merge with main progenitor 
represent surviving subhalos at $z=0$.
  \emph{Left:} A ``typical'' merger history, with a merger
of mass $m \simeq 0.1 M_0  \simeq 0.5 M_z$ at $z = 0.51$.
\emph{Right:} A halo that experiences an unusually
 large merger $m \simeq 0.4 M_0 \simeq 1.0 M_z$ at $z = 0.65$.}  
\label{mergertrees} \end{center}
\end{figure*}

\section{Simulations}
\label{Simulation}

Our simulation consists of $512^{3}$ particles, each  with mass $m_p =
3.16 \times 10^8 \Msun$, evolved within a comoving cubic volume of $80
h^{-1}$   Mpc on  a side using    the Adaptive  Refinement  Tree (ART)
$N$-body code \citep{Kravtsov97,Kravtsov04a}.  The cosmology is a flat
$\Lambda$CDM             model,          with               parameters
$\Omega_{M}=1-\Omega_{\Lambda}=0.3$, $h=0.7$,  and   $\sigma_{8}=0.9$.
The simulation root   computational  grid consists of  $512^3$  cells,
which are  adaptively refined to  a maximum of eight levels, resulting
in a  peak spatial resolution of  $1.2 h^{-1}$ kpc, in
comoving units.  This simulation  and the methods  we use to construct
merger trees have been discussed elsewhere
\citep{allgood06,wechsler06}.  Here we give a brief overview and refer
the reader to those papers for a more complete discussion.

Field dark matter halos and subhalos are identified using a variant of
the bound density maxima algorithm \citep{Klypin99}.  A \emph{subhalo}
is defined as a dark matter halo whose center is positioned within the
virial radius of another, more massive halo.  Conversely, we define a
\emph{field halo} to be a dark matter halo that does not lie within
the virial radius of a larger halo.  The virial radius is defined as the radius
of a collapsed self gravitating dark matter halo within which the
average density is $\Delta_{vir}$ times the mean density of the
universe.  For the family of flat cosmologies ($\Omega_m +
\Omega_\Lambda = 1$) the value of $\Delta_{vir}$ can be approximated
by \citep{Bryan&Norman98}:
\begin{equation}
\Delta_{vir} = \frac{18\pi^2 + 82(\Omega_m(z) - 1) - 39(\Omega_m(z)-1)^2}{\Omega_m(z)}.
\end{equation}
Masses,   $M$, are defined  for
field halos  as the mass enclosed within the virial radius, 
so that $M = (4\pi/3) R^3 \Omega_m \rho_c \Delta_{vir}$.
With these definitions, 
the virial radius for halos of mass $M$ at $z=0$ is given by:
\begin{equation}
R \simeq 205 h^{-1} \mbox{kpc } \left(\frac{M}{10^{12} \Msun}\right)^{1/3}.
\label{RvsM}
\end{equation} 
Note that this mass definition based on a fixed overdensity is largely
conventional, and is traditionally used as a rough approximation for
the radius within which the halos are virialized.  We refer the reader
to the recent work of \cite{Cuesta07} for further discussion of this
issue.  

Halo masses become more difficult to define in crowded environments.   
For example, if two halos are located within two virial radii of each  
other,  mass double-counting can become a problem. Also, subhalos can  
become tidally stripped if they are accreted into a larger halo.  
While the stripped material typically remains bound to the 
larger host halo, it is no longer bound to the smaller subhalo and should not be included 
in the subhalo's mass.  In  
these cases, the standard virial over-density definitions are not  
appropriate.    In order to overcome this ambiguity, we always define  
a halo's radius and mass as the minimum of the
virial mass and a ``truncation mass'' -- defined as the mass within the  
radius where the log-slope of the halo density profile becomes  
flatter than $-0.5$.  This definition of truncation mass is a relatively 
standard pratice when dealing with simulations of this kind 
\citep[e.g.][]{KlypinHoltzman97,Kravtsov04a,z05}, and
we follow this convention to remain consistent with other work in this field. 
In practice, our field  halos have masses and  
radii defined by the standard virial relations ($\sim 98 \%$ of all non- 
subhalos).
It is fair then to interpret our merger rates as infall rates into  
halo virial radii.   The masses
of objects just prior to infall are more likely affected by this  
definition, but the overall effect
on our results is not large.  As a test, we have redone our main analysis  
using
an (extrapolated) virial mass for infalling halos.  The results on  
fractional merger rates change only at the $\sim 5 \%$ level.

In the event that a halo experiences a close pass with another halo---entering
within the virial radius for a short time, then exiting the virial radius
never to return---the two halos are considered isolated, even though one 
may lie within the virial radius of the other.
Conversely, if the smaller halo falls back within the virial radius and the two halos
subsequently merge together, then we continue to consider the smaller halo a subhalo
even during the time when it lies outside the virial radius of its host.  
This has been referred to as the
``stitching'' method---as opposed to ``snipping,'' which would count
the above example as two separate mergers \citep{FakhouriMa07}.

By constructing mass functions, we find that these halo catalogs
are complete to a minimum mass $10^{10} \Msun$.  This allows us to measure
the accretion of objects $10$ times smaller than $10^{11} \Msun$ halos,
or objects up to $1000$ times smaller than $10^{13} \Msun$ halos.
In all cases, we denote the mass of an accreted object as $m$.
Overall we have a total of 17,241 halos in our sample with $M_0 >
10^{11} \Msun$, and (6642, 2479, 911, and 298) halos in
logarithmically-spaced mass bins centered on $\log M_0 =$ ($11.5$,
$12.0$, $12.5$, and $13.0$) respectively, in units of $\Msun$.

Our merger tree construction mirrors that described in
\cite{Kravtsov04a} and uses 48 stored timesteps that are approximately
equally spaced in expansion factor between the current epoch $a =
(1+z)^{-1} = 1.0$ and $a = 0.0443$.  We use standard terminologies for
\emph{progenitors} and \emph{descendant}.  Any halo at any timestep
may have any number of \emph{progenitors}, but a halo may have only
one \emph{descendant} --- defined to be the single halo in the next
timestep that contains the majority of this halo's mass.  We use the
terms ``merger'' and ``accretion'' interchangeably to designate the
infall of a smaller halo into the virial radius of a larger one.  The
term {\em main progenitor} is used to reference the most massive
progenitor of a $z=0$ halo tracked continuously back in time.

Throughout most of this work we present results in terms of absolute
mass thresholds on the infalling mass $m$.  Our principle statistics
are quantified using the infalling mass thresholds in terms of the
final $z=0$ mass of the main progenitor halo (e.g.  $m > 0.1 M_0$).
Indeed, many of our results are approximately self-similar with
respect to halo mass when the infalling mass cut is defined in this
scaled manner.  By definition, the maximum mass that a merging halo
can have is $m = 0.5 M_0$.  Note that it is common in the literature
to study the {\em merger ratio} of an infalling object, $ m/M_z$,
where $M_z$ is the main progenitor mass at the redshift $z$, just
prior to the merger.  Here, $M_z$ {\em does not} incorporate the mass
$m$ itself and therefore $m/M_z$ has a maximum value of $1.0$.  A
parallel discussion that uses $m/M_z$ is presented in the Appendix,
but the absolute mass thresholds are used as our primary means to
quantify merger statistics in the main part of this paper.  We make
this choice for two reasons.  First, it is relatively easy to
understand completeness effects using a fixed threshold in $m$, while
completeness in $m/M_z$ will vary as a function of time and will
change from halo to halo depending on its particular mass accretion
history. Second, an event with $m/M_z \sim 1$ does not necessarily
imply that the infalling object $m$ is large compared to the final
halo mass $M_0$.  In order to be conservative, we would like to
restrict ourselves to mergers that are large in an absolute sense
compared to typical galaxy masses today.

Figure \ref{mergertrees} shows two pictorial examples of merger trees
for halos with approximately equal $z=0$ masses $M_0 \simeq 10^{12.5}
\Msun$.  Time runs from top to bottom and the corresponding redshift
for each timestep is shown to the left of each tree.  The radii of the
circles are proportional to the halo radius $R \sim M^{1/3}$, while
the lines show the descendent--progenitor relationship.  The color and
type of the connecting lines indicate whether the progenitor halo is a
field halo (solid black) or a subhalo (dashed red).  The most massive
progenitor at each timestep --- the main progenitor --- is plotted in
bold down the middle.  The ordering of progenitor halos in the
horizontal direction is arbitrary.  Once a halo falls within the
radius of another halo, it becomes a subhalo and its line-type changes
from black solid to red dashed.  When subhalo lines connect to a black
line this corresponds to a central subhalo merger or to a case when
the subhalo has been stripped to the point where it is no longer
identified.  When field halos connect directly to a progenitor without
becoming subhalos in the tree diagram it means that the subhalo is
stripped or merged within the timestep resolution of the simulation.
Halos that are identified as subhalos of the main halo at $z=0$ are
represented by the dashed-red lines that reach the bottom of the
diagram without connecting to the main progenitor line.

Note that the extent to which we can track a halo after it has become
a subhalo, and the point at which a subhalo is considered ``destroyed'' is dependent
both on spacing of our output epochs and mass resolution of the simulation.  This
is another reason why we count mergers when a halo falls within the virial
radius (when the lines in Figure 1 change from solid-black to dashed-red) and not
when a subhalo experiences a central merger with its host.

The left diagram (``halo 810'') in Figure 1 shows a fairly typical
merger history, with a merger of mass $m \simeq 0.1 M_0$ at $z \simeq
0.51$.  The merger ratio at the time of the merger was $m/M_z \simeq
0.5$.  The right diagram (``halo 853'') shows a very rare type of
merger history with a massive event $m \simeq 0.4 M_0$ at $z \simeq
0.65$. This was a nearly equal-mass accretion event at the time of the
merger, $m/M_z \simeq 1.0$.  Note that neither of these large mergers
survive for long as resolved subhalos --- they quickly lose mass and
merge with the central halo.  Each of these halos has two $\sim
10^{10} \Msun$ subhalos that survive at $z=0$.


\section{Results}
\label{Results}

\begin{figure}[t!]
\includegraphics[width=0.45\textwidth]{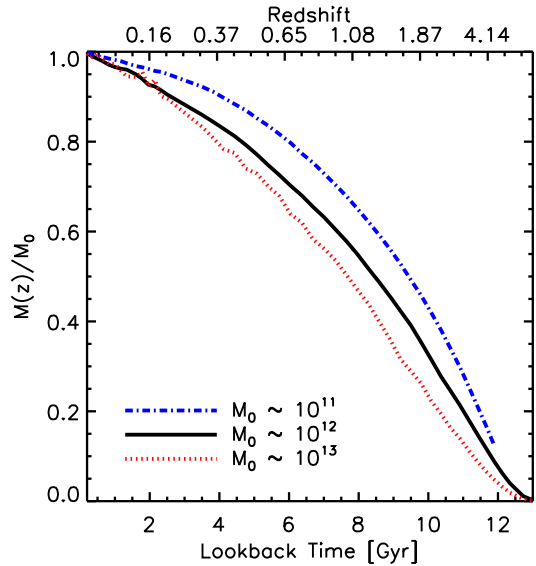} 
\caption{Average mass accretion histories for three bins in halo mass,
  as a function of lookback time.  Each bin gives the average for bins
  of size log$M_0$ = 0.5, centered on the stated value.  More massive
  halos accreted a larger fraction of their mass at late times.}
\label{Mvsz}
\end{figure}

\subsection{Accretion Histories and Mass Functions}

The literature is rich with work on the cumulative mass accretion
histories of halos as a function of redshift \citep[e.g][and
references therein]{wechsler02,zhao03,tasitsiomi04,li07}.  We begin by
re-examining this topic for the sake of completeness.  Figure
\ref{Mvsz} shows average main progenitor mass accretion histories,
$M_z = M(z)$, for halos of three characteristic final masses, $M_0 =
M(z=0)$.  We confirm previous results that halo mass accretion
histories are characterized by an initial rapid accretion phase
followed by a slower accretion phase, and that more massive halos
experience the rapid accretion phase later than less massive halos
\citep[][]{wechsler02}.  Milky Way-sized halos with $M_0 = 10^{12}
\Msun$ will, on average, accrete half of their mass by $z \simeq 1.3$,
corresponding to a lookback time of $\sim 8.6$ Gyr.

\begin{figure}[t!]
\includegraphics[width=0.485\textwidth]{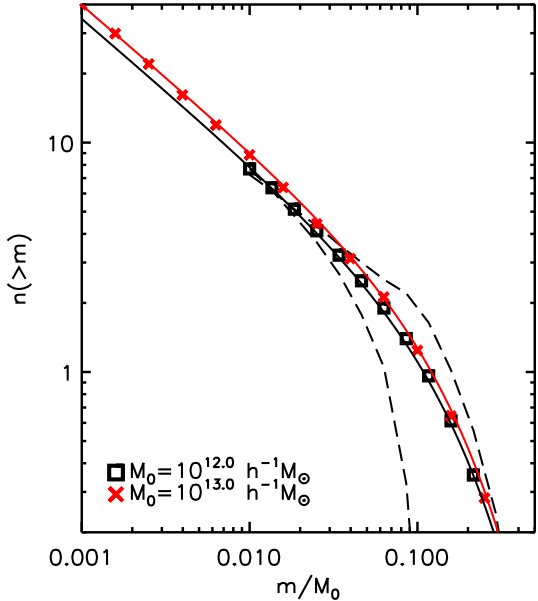} 
\caption{Mass functions of accreted material, with respect to 
  the final halo mass.  Lines show the cumulative
 number of  mergers that a halo experiences 
 with objects larger than   $m/M_0$, integrated over the
main progenitor's formation history.  
The (black) squares show the average for $10^{12} \Msun$ halos; (red)
crosses show the average for $10^{13} \Msun$ halos.  Lines through the
data points show the fits given by Equation \ref{eq:ngtm}.  The upper/lower dashed
lines indicate the $\sim 25 \% / 20 \%$ of halos in the $10^{12} \Msun$
sample that have experienced exactly two/zero $m \geq 0.1 M_0$ merger
events.  Approximately $45 \%$ of halos have exactly one $m \geq 0.1
M_0$ merger event; these systems have mass accretion functions that
resemble very closely the average.}
\label{NgtM}
\end{figure}

\begin{figure}[t!]
    \includegraphics[width=.49\textwidth]{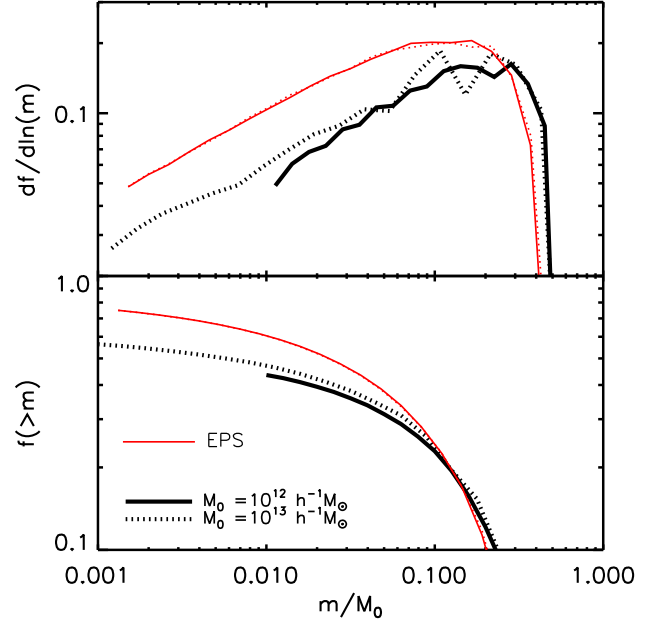} 
\caption{ The fraction of halo's final mass $M_0$ 
accreted from objects of mass $m/M_0$,
integrated over time.  The
differential (upper) and cumulative (lower)
mass fractions are shown. The thick (black) 
lines show simulation results and the
thin (red) lines show the prediction from standard 
Extended Press Schechter (EPS).
Accretions with
 $m \simeq (0.03 - 0.3) M_0$ dominate the mass buildup in halos of all $M_0$.
The distributions are approximately self-similar with host halo mass.
This is in reasonably good agreement with the EPS expectation, although
the EPS fractions systematically sit above the simulation results.
}
\label{dfdm}
\end{figure}

\begin{figure*}[tr]
\begin{center}
    \subfigure{\includegraphics[width=.45\textwidth]{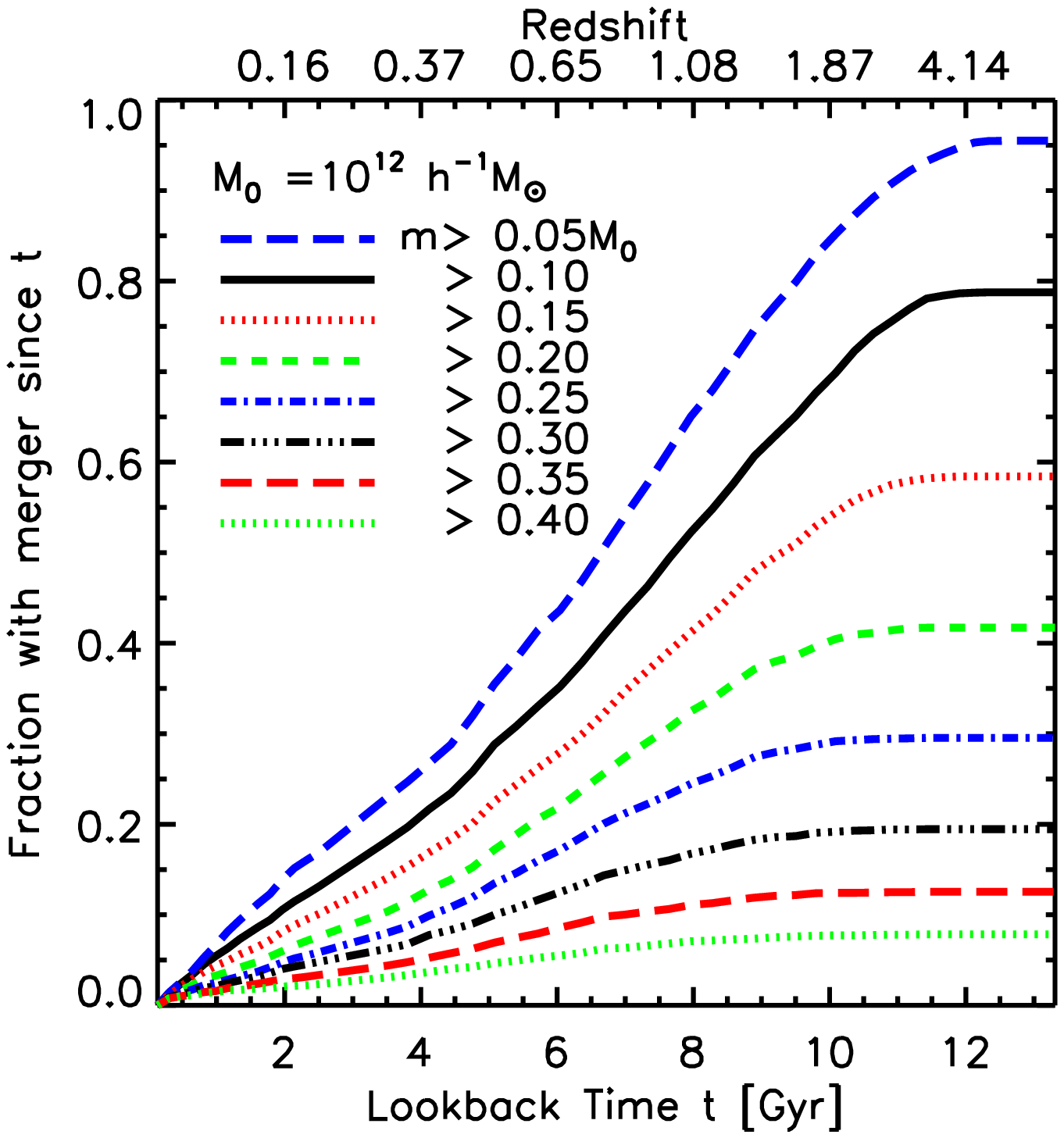}} 
    \subfigure{\includegraphics[width=.45\textwidth]{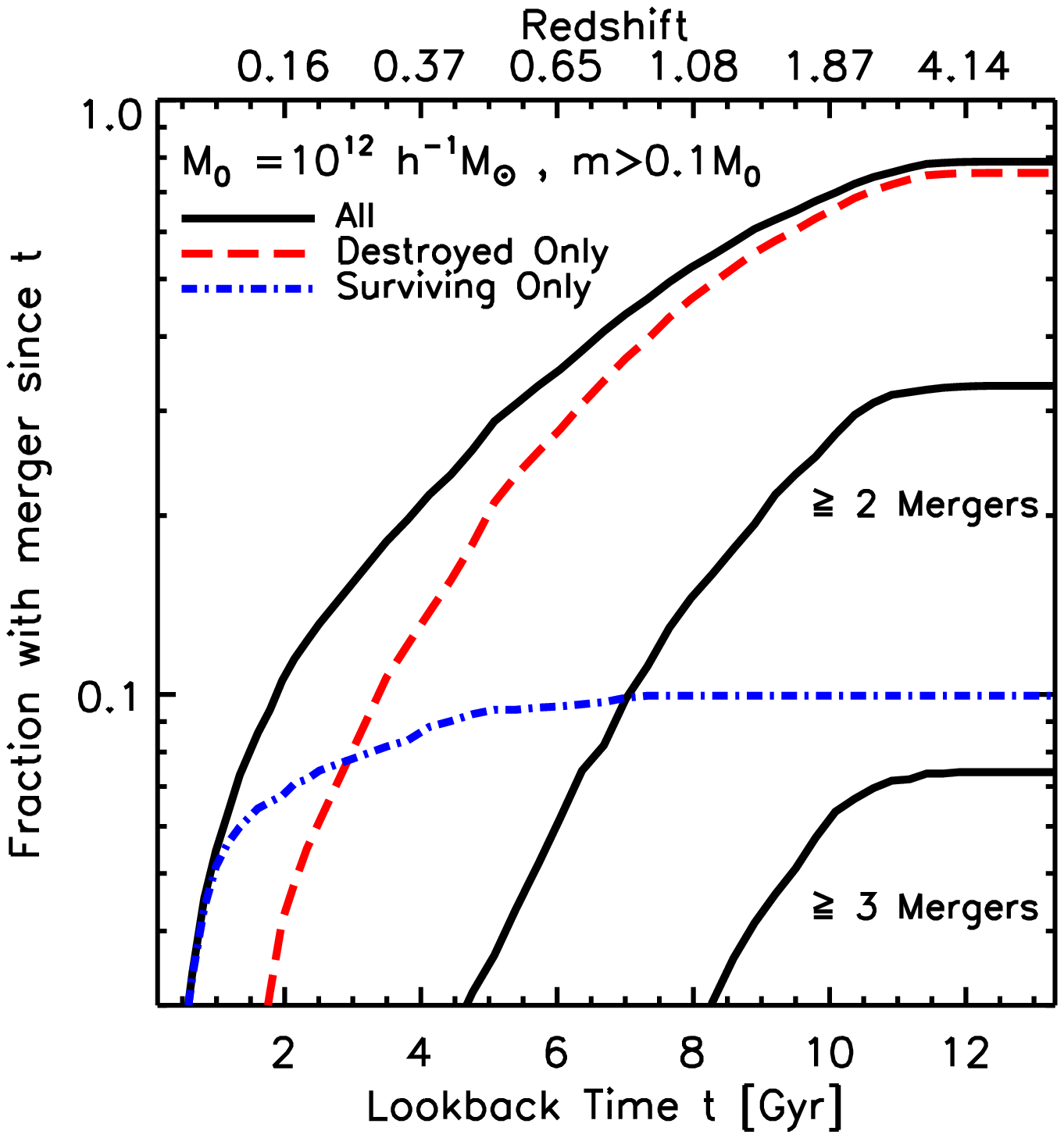}} 
    \caption{{\em Left:} The fraction of Milky Way-sized halos, $M_0
      \simeq 10^{12} \Msun$, that have experienced at least one merger
      larger than a given mass threshold, $m$, since look-back time t.
      {\em Right:} The three solid (black) lines show the fraction of
      $M_0 \simeq 10^{12} \Msun$ halos that have experienced at least
      one, two, or three mergers larger than $m = 0.1 M_0$ in a
      lookback time $t$.  The upper solid line is the same as the
      solid line in the left panel.  The dashed (red) line include
      only mergers that are ``destroyed'' before $z=0$ ({\em i.e.}
      lose $> 80 \%$ of their mass by $z=0$).  Conversely, the
      dot-dashed (blue) line shows the fraction of halos with at least
      one merger that ``survives'' as a subhalo as a function of
      lookback time to the surviving halo's merger.  Large accretions
      of $m > 0.1 M_0$ typically do not survive for more than $\sim 3$
      Gyr after accretion.}
\label{MMfrac} 
\end{center}
\end{figure*}

While Figure \ref{Mvsz} provides some insight into {\em when} mass is
accreted into halos, we are also interested in characterizing {\em
  how} this mass is accreted. (Ultimately, we will present merger statistics 
for a joint distribution of both time {\em and} mass ratio.)  
Now we investigate the mass function
$n(m)$ of objects larger than $m$ that have merged into the main
progenitor over its history.  The solid line in Figure \ref{NgtM}
shows $n(m)$ averaged over halos in the $M_0 = 10^{12} \Msun$ bin,
plotted as a function of $m/M_0$.  On average, Milky Way-sized halos
with $M_0 \simeq 10^{12} \Msun$ experience $\sim 1$ merger with
objects larger than $m \sim 10^{11} \Msun$, and $\sim 7$ mergers with
objects larger than $m \sim 10^{10} \Msun$ over the course of their
lives.

For some purposes, an analytic characterization of the accreted mass
function will be useful.  We have investigated the average $n(m)$
function for halos in the mass range $M_0 = 10^{11.5} - 10^{13} \Msun$
and find that the shape of this function is remarkably similar over
this range (smaller $M_0$ were neglected in order to achieve a
reasonable range in $m/M_0$).  Specifically, we find that $n(m)$ is
well-characterized by a simple function of $x \equiv m/M_0$:
\begin{equation}
n(>x) = A \, x^{-\alpha} \, (0.5-x)^{\beta}, 
\label{eq:ngtm}
\end{equation}
with $\beta = 2.3$, $\alpha = 0.61$, and
 $x \leq 0.5$ by construction.
Interestingly, we find that the overall normalization increases
monotonically with halo mass, and that the trend can be
approximated  as 
$A(M_0) \simeq 0.47 \log_{10}(M_0) -3.2$, where $M_0$ is in units of $\Msun$.
This mass-dependent normalization, together with Equation \ref{eq:ngtm}, 
reproduces our measured $n(m)$ functions quite well --- to better than
$5 \%$ at all $m$ in smaller halos
($M_0 = 10^{11.5} - 10^{12} \Msun$), and (somewhat worse)
to $15 \%$ at higher masses ($M_0 = 10^{12.5} - 10^{13} \Msun$).
\footnote{The quoted errors are restricted to
$n > 0.05$.}
  
We would also like to understand the scatter in the accreted mass
function from halo to halo at fixed $M_0$.  It is not appropriate to
simply describe the variation in $n(m)$ at a fixed mass, because the
total mass accreted is constrained to integrate to less than $M_0$.
This means that the number of small objects accreted may be
anti-correlated with the number of large objects accreted.  With this
in mind, we provide an illustration of the scatter with the two dashed
lines in Figure \ref{NgtM}.  The upper dashed line shows the average
$n(m)$ for the $\sim 25 \%$ of halos that have experienced exactly two
accretion events larger than $m = 0.1 M_0$.  The lower dashed line
shows the average $n(m)$ for the $\sim 20 \%$ of halos that have
experienced exactly zero $m > 0.1 M_0$ accretion events. Approximately
$\sim 45 \%$ of halos have exactly one $m > 0.1 M_0$ event, and these
have an average accreted mass function that is very similar to the
overall average shown by solid (black) line in Figure \ref{NgtM}.
Halos with fewer large mergers show a slight tendency to have more
small mergers, but the effect is not large.

Figure \ref{dfdm} presents some of the same information shown in
Figure \ref{NgtM}, but now in terms of the mass fraction, $f(m)$,
accreted in objects larger than $m$ for $M_0 = 10^{12}$ and $10^{13}
\Msun$ halos (thick lines, see legend).  The upper panel in Figure
\ref{NgtM} shows the differential fraction, ${\rm d} f/{\rm d}\ln m =
(-m^2/M_0) {\rm d}n/{\rm d}m$, while the lower panel plots the
integrated fraction $f(m)$.  As before, we have normalized the
accreted masses, $m$, by the final $z=0$ main progenitor mass $M_0$.
We find that $f(>m)$ is also well fit by
Equation \ref{eq:ngtm} (to better than $10\%$ across all masses $M_0 = 
10^{12} - 10^{14} \Msun$)\footnote{The quoted errors are restricted to
$f > 0.1$.}.  As before, $x \equiv m/M_0$, but now 
$A(M_0) \simeq 0.17 \log_{10}(M_0)-0.36$, with $M_0$ still in units of $\Msun$.
The best fit parameters are $\alpha = 0.05$, and $\beta = 2.3$.
The lines are truncated at $m/M_0 = 0.01$ and $0.001$, corresponding
to our fixed resolution limit at $m = 10^{10} \Msun$.  The thin (red)
lines of the same line types show the same quantities predicted from
Extended Press Schechter \citep[EPS][]{lc93} Monte-Carlo merger trees.
Each of these lines is based on $5000$ trees generated using the
\cite{sk99} algorithm.

In broad terms, the mass spectrum of accreted objects agrees fairly
well with the EPS expectations, especially considering the relative
ambiguity associated with defining halo masses in simulations
\citep[e.g.][]{cw07,diemand07, Cuesta07}. However, it is worth
discussing the similarities and differences in some detail.  It is a
well-known expectation from EPS that the total mass accreted into a
halo of mass $M_0$ is dominated by objects of mass $m \sim 0.1 M_0$
\citep{lc93,zb03,purcell07,zentner07}.  Our simulations reveal that
indeed $m \simeq (0.03-0.3) M_0$ objects are the most important
contributors to the final halo mass.

EPS trees predict self-similar mass fractions across all halo masses.
Our more massive halos, however, show a slight tendency to have more
of their mass accreted in collapsed objects, across all scaled masses
$m/M_0$.  As discussed in association with Equation \ref{eq:ngtm}
above, the overall normalization of the mass spectrum is slightly
higher for our more massive halos.  For example, the mass accreted in
objects larger than $m = 0.01 M_0$ is $\sim 45 \%$ for $M_0 = 10^{12}$
halos and $\sim 50 \%$ for $M_0 = 10^{13}$ halos.  Both of these
fractions are low compared to the $\sim 65 \%$ expected from the
semi-analytic EPS merger trees.

It is interesting to estimate  the {\em total} mass fraction  accreted
in collapsed objects ($m>0$) by extrapolating  the $n(m)$ fit given in
Equation \ref{eq:ngtm} to $m \rightarrow 0$.  We find
\begin{eqnarray}
  f(>0) & = & \int_0^{0.5 M_0}  \frac{{\rm d}f}{{\rm
d}m}    \,  {\rm   d}m   \nonumber \\
	& = & \int_0^{0.5} A \, x^{1-\alpha} \, (0.5-x)^{\beta} \, {\rm d}x \\ 
& \simeq & 0.24 \, A . \nonumber
\end{eqnarray}
In the second step, $\beta$ and
$\alpha$ are the shape parameters in Equation \ref{eq:ngtm}.  In the
last step we have used our best fit parameters $\alpha = 0.61$ and
$\beta = 2.3$.  Using Equation 2 for $A(M_0)$, we find that the total
accreted mass fraction (in virialized halos) increases from $f \simeq
0.50$ to $0.70$ as $M_0$ varies from $10^{11.5}$ to $10^{13} \Msun$.
This suggests that a significant fraction ($\sim 30-50 \%$) of dark
halo mass is accreted in the form of ``diffuse'', unvirialized
material, and that smaller halos have a higher fraction of their mass
accreted in this diffuse form.  Of course, our conclusion on
``diffuse'' accretion may be due, at least in part, to difficulties in
precisely defining halo masses---and, in particular, in defining halo
masses in dense environments where mergers are occurring rapidly.
It is  also  possible  that  the accreted  mass  function
 steepens below  our resolution limit, resulting in a lower diffuse 
fraction than we expect from our extrapolation, 
but there is no clear physical reason to expect such a steepening.

\begin{figure*}[htp]
  \begin{center}
    \subfigure{\includegraphics[width=.45\textwidth]{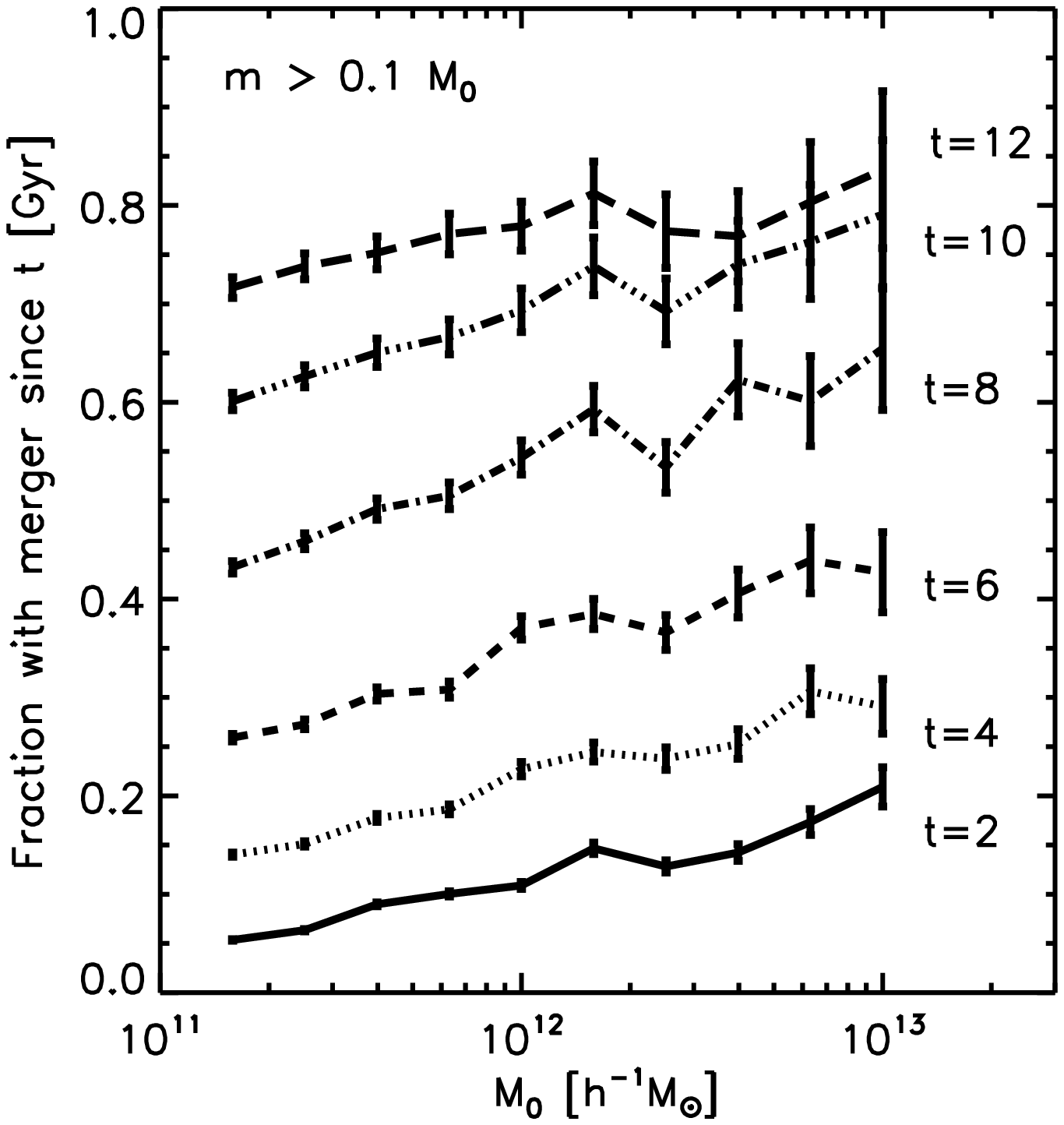}} 
    \subfigure{\includegraphics[width=.45\textwidth]{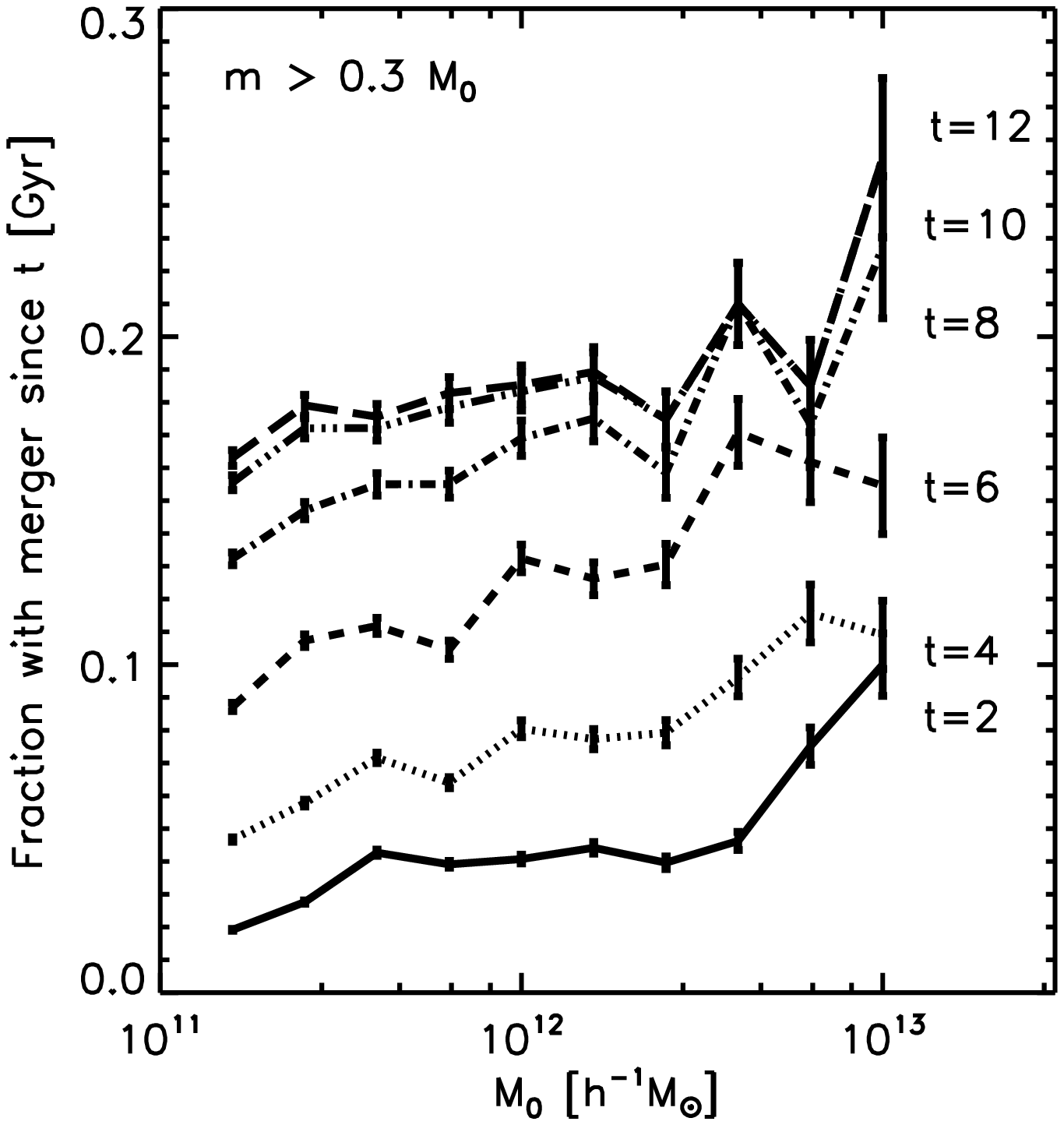}} 
    \caption{
The fraction of halos of mass $M_0$ at $z=0$ 
that have experienced a merger with an object more massive than
$0.1 M_0$ (left) and $0.3 M_0$ (right) in the last $t$ Gyr. 
Error  bars are Poissonian based on the number of halos used in each
mass bin.   Note the  fairly weak mass dependence.}
\label{LMsincet} \end{center}
\end{figure*}


\subsection{Merger Statistics}

Understanding how galaxy mergers can affect
 galaxy transformations and morphological fractions
necessarily requires an understanding of halo merger statistics. 
Of specific interest is the overall fraction of halos that have
had mergers within a given look back time.

The left panel of Figure \ref{MMfrac} shows the fraction of Milky
Way-sized halos ($M_0 \sim 10^{12} \Msun$) that have experienced {\em
  at least one} ``large'' merger within the last $t$ Gyr.  The
different line types correspond to different absolute mass cuts on the
accreted halo, from $m > 0.05 \, M_0$ to $m > 0.4 \, M_0$.  The
tendency for lines to flatten at large lookback times is a physical
effect and {\em is not an artifact of limited resolution}.
Specifically, the lines flatten at high $z$ because the halo main
progenitor masses, $M_z$, become smaller than the mass threshold on
$m$ (see Figure 2).  We find that while less than $\sim 10 \%$ of Milky
Way-sized halos have {\em ever} experienced a merger with an object
large enough to host a sizeable disk galaxy, ($m > 0.4 \, M_0 \simeq 4
\times 10^{11} \Msun$), an overwhelming majority ($\sim 95 \%$) have
accreted an object more massive than the Milky Way's disk ($m > 0.05
M_0 \simeq 5 \times 10^{10} \Msun$).  Approximately $70 \%$ of halos
have accreted an object larger than $m \simeq 10^{11} \Msun$ in the
last $10$ Gyr.

While the left panel of Figure \ref{MMfrac} illustrates the fraction
of halos with at least one merger in a given time, the right panel of
Figure \ref{MMfrac} provides statistics for multiple mergers.  We
again focus on $M_0 \sim 10^{12} \Msun$ halos, but restrict our
statistics to accretions with $m > 0.1 \, M_0 \simeq 10^{11}
\Msun$. The upper solid (black) line shows the fraction of halos with
at least one accretion event within the last $t$ Gyr.  This reproduces
the solid (black) line in the left-hand panel, but now the vertical
scale is logarithmic.  The middle and lower solid (black) lines in the
right panel show the fraction of halos with at least two and at least
three mergers larger than $0.1 \, M_0$ in the past $t$ Gyr,
respectively.  Roughly $\sim 30 \%$ of Milky Way-sized halos have
experienced at least two accretion events larger than $\sim 10^{11}
\Msun$.  Multiple events of this kind could be important in forming
elliptical galaxies \citep{hernquist93,bk05,robertson06b,naab06}.

As discussed above, our merger events are defined at the time when the
accreted halos first cross within the virial radius of the main
progenitor.  This definition allows us to focus on a robust statistic
that is less likely to be affected by baryonic components.  In \S
\ref{Discussion}, we speculate on the implications of these results for
disk stability.  In this context, one may be concerned that some
fraction of our identified ``mergers'' will never interact with the central disk
region, but instead remain bound as surviving halo substructure.  We
expect this effect to be most important for smaller accretions, as
larger merger events will decay very quickly.

The dashed (red) line in Figure \ref{MMfrac} shows a statistic that is
analogous to the upper black line --- the fraction of halos that have
had at least one merger in the last $t$ Gyr --- except we have now
restricted the analysis to include only objects that are ``destroyed''
before $z=0$.  We define an object to be ``destroyed'' if it loses
more than $80 \%$ of the mass it had at accretion because of
interactions with the central halo potential.  (Our results are
unchanged if we use $70-90\%$ thresholds for mass loss).  Likewise,
the dot-dashed (blue) line shows the same statistic restricted to
``surviving'' objects.  Only $\sim 10 \%$ of Milky Way-sized halos
have surviving massive substructures at $z=0$ that are remnants of $m
\simeq 10^{11} \Msun$ accretion events.  These survivors were
typically accreted within the last $\sim 3$ Gyr.

We conclude that large accretions that happened more than $\sim 3$ Gyr
ago would have had significant interactions with the central disk
regions of the main progenitor.  Indeed, this must be the case if the
Milky Way is ``typical''.  The most massive Milky Way satellite, the
LMC, was likely accreted with a mass no larger than $m \sim 4 \times
10^{10} \Msun$ \citep{vandermarel02,robertson05}.  According to Figure
\ref{NgtM}, we expect that the Milky Way has accreted at least $\sim
3$ objects that are larger than the LMC over its history.  The
expectation is that most of the larger objects that were accreted have
been shredded by the central galaxy potential, and have deposited
their stars in the extended stellar halo \citep{bj05}.  Interestingly,
a recent re-analysis of the LMC's motion suggests that it is indeed on
its first passage about the Milky Way \citep{besla07}, as we would
expect for surviving, massive satellites.

Given  that  galaxy  morphology  fractions   are observed  to   change
systematically     as  a   function  of   luminosity    or mass  scale
\citep[e.g.][]{Park07}, we are   also interested in  exploring  merger
fractions  as a function  of $M_0$.   The  left panel of  Figure
\ref{LMsincet} shows  the fraction  of  halos that have had  a  merger
larger than $m = 0.1 \, M_0$ within the last $t  = 2$, $4$, ... $12$ Gyr,
as indicated by the $t$ labels.  The  right panel shows the same
statistic computed  for  larger $m > 0.3 \, M_0$  accretion   events.  The most
striking result is that  the merger fraction  is fairly independent of
$M_0$.  This suggests that halo merger statistics alone cannot explain
the tendency   for early-type spheroidal   galaxies to reside  in more
massive halos.  Baryon physics must instead play the primary role in 
setting this trend.  (The right-panel in Figure
\ref{LMsincet} does show evidence that  halos larger than $M_0 \sim  3
\times 10^{12} \Msun$ have a higher fraction of recent ($t \lesssim 2$
Gyr) large ($m > 0.3 M_0$) events, but the  overall merger fraction is
small ($\lesssim 10 \%$), even  for group-sized halos with $M_0 \simeq
10^{13} \Msun$.)  When counting {\em instantaneous} merger rates there
appears to be  a relatively universal merger rate  across a wide range
of host  masses  $M_0  \sim  10^{12}$--$10^{14}$ \citep{FakhouriMa07}.
Given  this uniformity, we  speculate  that the  weak trend  we see in
Figure \ref{LMsincet} may  ultimately be a direct  result of the trend
for more massive halos to form later, as shown in Figure \ref{Mvsz}.

If a merger occurs  early enough, we might  expect the main progenitor
halo to grow significantly   after the merger.   The fraction  of mass
accreted  since  the  last merger  is potentially   important, as, for
example, it could enable the regrowth of  a destroyed disk.  One could
reason that  the fraction of halos  that experience a large merger and
then subsequently fail to accrete a  significant amount of mass is the
most  relevant  statistic  for evaluating   the   probability of  disk
formation.  However, we find that in most cases, very little mass
is accreted after an event that is large relative to $M_0$.

Figure \ref{Macc} shows the average {\em fraction} ($\Delta M$/$M_0$)
of a halo's final mass $M_0$ that is accreted since the last large
merger.  Each curve corresponds to a different threshold in $m$.  As
in Figure \ref{LMsincet}, the overall trend with final mass $M_0$ is
very weak.  We define the accreted mass by $\Delta M = M_{0} - M_{\rm
  m}$, where $M_{\rm m}$ is the mass of the host halo's main
progenitor {\em after} the most recent large merger.  The upper
(solid, black) line includes all halos that have had at least one
event larger than $m = 0.1 \, M_0$ and the lowest line (long-dashed,
red) includes all halos that have had at least one merger larger than
$m = 0.35 \, M_0$.  This shows that the fractional mass accreted since
a merger is a decreasing function of $m/M_0$.  It also shows that
the fraction is typically small, with $\Delta M/M_0 \lesssim 30 \%$
($10 \%$) for $m \gtrsim 0.1 \, M_0$ ($0.3 \, M_0$) mergers.  These
trends are a consequence of the fact that we are doing this
calculation at a fixed $M_0$ --- if a merger $m$ is large compared to
$M_0$, then there is little room in the mass budget for new material
to be accreted after the merger.  This implies, for example, that if a
disk is destroyed as a result of a large-$m/M_0$ accretion event, it
is unlikely that a new ``disk-dominated'' system can be regrown from
material that is accreted into the host halo after the merger.
However, gaseous material involved in the merger may re-form a disk 
\citep[see e.g.][]{ZQS88, robertson06b}.

\section{Discussion}
\label{Discussion}

\subsection{Milky Way Comparison}
The Milky Way has a dark matter halo of mass $M_0 \simeq 10^{12} \Msun$
\citep{klypin02}, and its stellar mass 
is dominated by a thin disk of mass $ \simeq 3.5
\times 10^{10}  M_\odot$  \citep{klypin02,widrow_dubinski05}.
The thin disk has  vertical scale height  that is just $\sim  10 \%$ of its
radial scale  length  \citep{Siegel02,Juric05,Newberg06}, and contains
stars as  old as  $\sim  10$ Gyr \citep{nordstrom04}.
 Moreover,
stars in the local thick disk are predominantly older than $\sim 10$ Gyr, 
and the bulge is old as well.  This suggests that there
was not significant merger activity in the Milky Way to drive
gas towards the bulge or to thicken the disk in the past
$\sim 10$ Gyr \citep[][]{wyse01}.

Based on our results, a galaxy like the Milky Way would seem rare in a
$\Lambda$CDM universe.  Roughly $70\%$ of dark matter halos of mass 
$M_0 \simeq 10^{12} \Msun$ have experienced a merger with a halo of mass 
$10^{11} \Msun$ in the past $\sim 10$ Gyr.  A merger of this size should thicken the 
existing disk and drive gas into the center of the galaxy to create a bulge \citep{k07}.  If the 
Milky Way has \emph{not} experienced a merger of this magnitude, that would make
our galaxy a rare occurrence ($\lesssim 30\%$ of halos).  On the other hand, if the Milky Way 
\emph{has} experienced such a merger, it is difficult to understand its observed 
early-type morphology and thin-disk properties.

\subsection{Morphological Fractions and Thick Disks}
  The
degree to which the Milky Way halo is typical for its mass is becoming
better   understood thanks to the   advent of large,  homogeneous
astronomical   sky  surveys.   As mentioned in the introduction,
broad-brush categorizations of  ``late type'' vs. ``early type''
suggest that $\sim 70 \%$ of Milky Way-sized halos host late-type galaxies
\citep[e.g.][]{Weinmann06,vandenBosch06,Ilbert06alt,Choi07,Park07}.
 The degree   to  which  ``late-type''   is synonymous   with   ``thin
 disk-dominated'' is difficult to quantify with current data sets, but
 for the sake  of this discussion, we will  assume that this is the
case.  Also discussed earlier were the results of
Kautsch et al. (2006), who
found that $\sim 16 \%$ of disk galaxies are bulgeless
systems.  This suggests that  $\sim (0.7)(0.16) \sim 11 \%$ of
Milky Way-sized halos host pure disk galaxies.

The observed morphological fractions may be compared to the halo
merger fractions presented in Figure \ref{MMfrac}.  These results show
that an overwhelming majority of Milky Way-sized halos ($\sim 95 \%$)
experience at least one merger larger than the {\em current} mass of
the Milky Way disk ($\gtrsim 5 \times 10^{10} \Msun$).  Figure
\ref{NgtM} shows that a typical $M_0 \simeq 10^{12} \Msun$ halo has
merged with $\sim 2-3$ objects of this size over its history.  It is
possible that mergers of this characteristic mass are responsible for
creating thick disk components in most galaxies
\citep{walker_etal96,dalcanton_bernstein02}.  More detailed
simulations will be required to test whether disks are destroyed or
overly thickened by the predicted infall of $m \sim 5 \times 10^{10}
\Msun$ objects, and whether these thickening events happen too late to
explain thick disks as old as those observed
\citep{dalcanton_bernstein02}.  Understanding how bulgeless galaxies
could exist in halos with mergers of this kind is a more difficult
puzzle.

\begin{figure}[t!]
   \includegraphics[width=.48\textwidth]{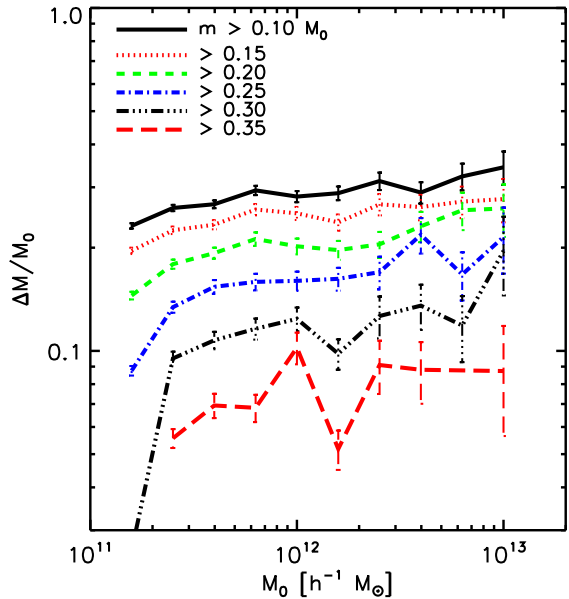} 
    \caption{The average fractional change in a halo's main progenitor mass $\Delta M/M_0$ 
     since its  last large merger, shown  as a  function of the halo's
     $z=0$ mass $M_0$.  Each line includes  only halos that have had a
     merger larger than the $m$ value indicated ({\em i.e.} only halos
     that have had a $m > 0.35 M_0$ accretion event are included in the
     lowest  line).   Error bars shown are  Poisson  on  the number of
     halos  used in the average.}
\label{Macc} \end{figure}

Perhaps more disturbing for the survival of thin disks are the
statistics of more substantial merger events.  Figure \ref{MMfrac}
shows that $m \gtrsim 10^{11} \Msun$ accretions are quite common in
Milky Way-sized halos, with $\sim 70 \%$ of $M_0 \simeq 10^{12} \Msun$
objects experiencing such a merger in the past $\sim 10$ Gyr.  Of
course, the impact that these events will have on a central disk will
depend on orbital properties, gas fractions, and star formation in the
merging systems.  Generally, however, a merger with an object $\sim 4$
times as massive as the Milky Way thin disk would seem problematic for
its survival.

We find that a small fraction of a halo's final mass is typically
accreted into the main progenitor subsequent to $m \sim 0.1 M_0$
mergers --- this suggests that the regrowth of a dominant disk from
material accreted {\em after} such a merger will be difficult (see
Figure \ref{Macc}).  We conclude that if $\sim 70 \%$ of Milky
Way-sized halos contain disk-dominated galaxies, and if the adopted
$\Lambda$CDM cosmology is the correct one, then mergers involving $m
\simeq 10^{11} \Msun$ objects must not result in the destruction of
galaxy disks.  This is a fairly conservative conclusion because if we
naively match the percentages of mergers with an early-type fraction
of $\sim 30 \%$, then Figure 5 suggest that the critical mass scale
for disk survival is significantly larger, $m \gtrsim 2 \times 10^{11}
\Msun$.  Specifically, mergers involving objects that are $\sim 5 $
times the current Milky Way disk mass must not (always) destroy disks.

We remind the reader that the lookback times depicted in Figures 5 and
6 correspond specifically to the times when infalling halos first fall
within the main progenitor's virial radius.  Our estimates suggest
that the corresponding central impacts should occur $\sim$3 Gyr later
for the mass ratios we consider.  Therefore, when we quote merger
fractions to a lookback time of $\sim$10 Gyr, this will correspond to
an actual impact $\sim$7 Gyr ago.  Of course, the infalling systems
will also lose mass as they fall towards the central galaxy.  As we
have emphasized, the detailed evolution of merging objects can only be
determined with focused simulations, and the outcome of the subsequent
mergers will depend on the baryonic components and orbital properties
of the systems involved.

Kazantzidis et al.  (2007) have performed  a focused N-body simulation
in order  to investigate the  morphological response of a  thin, Milky
Way type stellar disk galaxy to a series of impacts with $6$ satellite
halos~\footnote{We note that their  $6$ accretions were chosen from  a
high-resolution  N-body simulation, and that this  number of events is
fairly  typical of what  we find based on  $\sim 2500$ Milky Way-sized
halos in Figure 4.   However, it is also typical  of a  Milky Way-size
halo  to  experience $1-2$ mergers   more  massive than any  of  these
accretions.}  of mass $m
\simeq (1   -  2) \times 10^{10}   M_\odot$
($\sim  30 - 60   \%$ of the disk  mass).   They find that  a dominant
``thin'' stellar disk component survives the bombardment, although its
scale-height was seen to increase from $250$ pc to  $\sim 400$ pc, and
a  second $\sim 1.5$   kpc ``thick'' component  was also  created.  In
addition, a new  central bar / bulge component  was also generated  in
these fairly small encounters.  While it is  encouraging to see that a
thin disk can survive some  bombardment, mergers with objects $\sim 6$
times  as massive as those considered  by Kazantzidis et al. should be
very common in Milky Way-sized  halos.  It remains  to be seen how the
infall of  $m \simeq    10^{11}   \Msun$  objects will   affect    the
morphologies of thin, $\sim 4 \times 10^{10} M_\odot$ stellar disks.  
Moreover, the merger history considered by Kazantzidis et al. extended to 
fairly recent events.  While there was no explicit star formation prescription
in these simulations, one would expect a broad range of stellar ages in the 
thickened disk stars in this case, instead of a predominantly old 
population---as seems to be observed in actual galaxies.

\subsection{Morphology--Luminosity Trends}

Another well-established observational trend is the
 morphology--luminosity        relation
\citep[recently,][]{Choi07,Park07}, which, when interpreted in   terms
of a morphology--halo mass relation,  demonstrates that the fraction of
late-type  galaxies   contained   within  dark    matter   halos    is
anti-correlated  with the mass  of  the halo \citep{Weinmann06}.   For
large galaxy halos, $M_0 \simeq 10^{13} \Msun$, the late-type fraction
is just $\sim 30 \%$, compared to $\sim 70 \%$ for  Milky Way-sized systems
\citep[][]{Weinmann06,vandenBosch06,Ilbert06alt,Choi07,Park07}.   
The result presented in Figure 6 is perhaps surprising in light
of this fact.  Specifically, merger
histories of galaxy halos are almost  self-similar in
$M_0$ when the infalling mass $m$ is  selected to be a fixed
fraction of $M_0$.  For example, $\sim 18 \%$ ($70 \%$) of $M_0 \simeq
10^{12} \Msun$ halos  have experienced an $m   > 0.3 M_0$ ($0.1  M_0$)
merger  event in    the  last $10$ Gyr.    This  fraction  grows  only
marginally to  $\sim 25 \%$ ($80  \%$) for $M_0  \simeq 10^{13} \Msun$
halos.  The  implication is  that dark  matter  halo  merger histories
alone cannot  explain  the  observed  correlation between   early-type
fraction and halo mass.  Specifically,   baryon physics must play  the
primary role in setting the observed trend between galaxy morphology
and halo mass.

\subsection{Successive Minor Mergers}

Recently, Bournaud et al. (2007) have used focused simulations to
investigate the response of a very massive $\sim 2 \times 10^{11}
M_\odot$ disk-dominated galaxy within a $\sim 10^{12} M_\odot$ halo to
mergers with total mass {\em ratios} ranging from $m/M_z = 0.02$ to
$1.0$.  Broadly speaking, they find that $m/M_z = 0.1$ merger ratio
events can transform their disk galaxy to an S0, and that $m/M_z =
0.3$ ratio events produce ellipticals.  It is unclear how these ratios
would change for a smaller primary disk mass, considering that
their disk mass is extremely massive for a halo of this size.  
In comparison, $\sim 95
\%$ of our Milky Way-sized halos experience an event with a merger
ratio of $m/M_z > 0.1$ (corresponding to $m \gtrsim 0.05 \, M_0$, see
Appendix) in the last 10 Gyr.  Similarly, $\sim 60 \%$ of our halos
experience $m/M_z > 0.3$ events (See Figure \ref{Afig1}). We note that
 the results of this type of simulation will be sensitive to the gas
fractions and ISM model of the interacting galaxies
\citep{robertson06b}.  

\cite{HayashiChiba06} have also investigated 
the response of a galactic disk to a succession of minor mergers of CDM subhalos.
They find that subhalos more massive than $15\%$ of the disk mass must not 
merge into the thin disk itself, or it will 
become thicker than the observed disk of the Milky Way.  While our merger rates are for 
subhalos entering the virial radius of the \emph{halo}, not when it penetrates the disk, 
we expect the fraction of disk mergers involving objects of mass $6 \times 10^9 M_{odot}$
to be quite high.  Recall that $95\%$ of halos experience an accretion event
larger than $5 \times 10^{10} M_{odot}$.  Even if these halos lose $90\%$ of
their mass before disk impact (which seems unlikely) they still meet the Hayashi and Chiba
criterion.  Note, however, that a detailed thin/thick decomposition may be 
required in order to fully evaluate this limit \citep{k07}.
A more detailed study of large mergers, 
including the necessary baryon physics, is required to fully explore this issue.

\subsection{Gas-rich Mergers}

Many of our results provide qualitative support to the idea that cool
gas-fractions play a fundamental role in governing the morphological
outcome of large mergers
\citep[][]{robertson06b,brook07b,brook07a,cox07}, with gas-rich
mergers essential to the formation and survival of disk galaxies.
While dark halo merger histories are approximately self-similar in
$M_0$, gas fractions are known to decrease systematically with halo
mass \citep[at least at $z=0$, see e.g.][and references
therein]{geha06,kaufmann07b}.  This implies that gas-rich mergers
should be more common in small halos than in large halos.  If gas-rich
mergers do allow for the formation or survival of disk galaxies, then
the gas-fraction-mass trend may provide an important ingredient in
explaining the observed morphology--mass trend.  Specifically, small
halos should experience more gas-rich mergers, while large halos
should experience more gas-poor mergers.  However, stars resulting 
from a gas-rich merger will be younger than the lookback time of the merger, 
suggesting that mergers of this type only serve as an adequate explanation
for early mergers.  Also, these gas-rich mergers would require 
sufficient angular momentum to keep the resulting disk from being 
too centrally concentrated.

\subsection{Comparison to Previous Work}

Semi-analytic  models  \citep[e.g.][]{diaferio01,okamoto01} and models
based   on    cosmological   SPH   simulations  \citep{maller06}   have
demonstrated  that  many  of   the   observed trends  between   galaxy
morphology,   density,   and    luminosity    can  be  explained     in
$\Lambda$CDM-based models.  However, these results rely on the ability
to {\em  choose} a characteristic  galaxy mass  ratio for morphological
transformation $r_{\rm char}$ (usually   $r_{\rm char} \sim 0.3$).   The
assumption is that that galaxy-galaxy mergers with $m_{\rm gal}/M_{\rm
gal} \lesssim  r_{\rm char}$ allow a  disk to remain intact, while all
larger mergers produce spheroids.
In a recent investigation, \cite{Koda07} have explored a two-parameter
model, where stellar spheroid formation depends on both $m/M_z$ and the absolute 
halo mass, $M_z$.  Host halos with masses smaller than a critical $M_z$ are assumed 
to have very low baryon content in this picture.  Using the PINOCCHIO density-field
algorithm \citep{Monaco02} to generate halo merger statistics, \cite{Koda07} find that the
fraction of disk-dominated galaxies can be explained if the only events that lead to 
central spheroid formation have $m/M_z > 0.3$ and $M_z \gtrsim 4 \times 10^{10} \Msun$.

The goal of this work has not been to find a ratio that can explain
morphological fractions.  Rather, our aim has been to emphasize the
relatively large number of more minor mergers that could potentially
be ruinous to disk survival.  As discussed in the Appendix, events
with $m \simeq 0.1 M_0$ in galaxy halos typically have $m/M_z \simeq
0.2$ at the time of accretion.  These $\sim 10^{11} \Msun$ mergers
would produce no morphological response in the primary disk under many
standard treatments \citep[e.g.][]{maller06,Koda07}.  
For the sake of comparison, Figure \ref{Afig1}
in the Appendix shows galaxy halo merger statistics for fixed $m/M_z$
cuts.  Our results are in good agreement with those derived by Koda et
al. (2007; Figure 4) using the PINOCCHIO algorithm, and with those
quoted by \cite{wyse04} for an analysis made by L. Hebb using the GIF
simulations.

Finally, we mention that \cite{cole07} have used the large Millennium
Simulation to investigate the progenitor mass functions of halos.  In
qualitative agreement with our findings, Cole et al. (2007) find that
the fraction of mass coming from halo progenitors is lower than
expected from standard EPS treatments.  Their approach was somewhat
different than ours, as they focused on the full progenitor mass
function, as opposed to the mass function of objects that merged into
the main progenitor as we have here.  They estimate that $\sim 14 \%$
of a halo's progenitors are not accounted for in collapsed objects at
any redshift.  This may be compared to our estimate of $\sim 30-50\%$
for the fraction of mass not {\rm directly} accreted in the form of
virialized objects for $M_0 = 10^{13}$ to $10^{12} \Msun$ halos.  The
differences between our numbers and theirs may come from the fact that
we have actually measured slightly different quantities.  We also used
different halo-finding algorithms and halo mass definitions, and
utilized different formulations to extrapolate the simulation results
to unresolved masses (a peak heights formulation in their case, and a
direct mass function formulation in our case).  A more thorough
investigation of the differences associated with halo-finding
algorithms and mass definitions is reserved for a future paper.  For
the purposes of this work, it is useful to point out that while a
direct comparison is difficult to make at this time, if anything, our
results on overall merger counts seem {\em low} compared to the
results given \cite{cole07}.

\section{Conclusion}

\label{Conclusion}

We have used a high-resolution $\Lambda$CDM $N$-body simulation to
investigate the merger histories of $\sim 17,000$ galaxy dark matter
halos with masses $M_0 = 10^{11-13} \Msun$ at $z=0$.  Mergers with
objects as small as $m = 10^{10} \Msun$ were tracked. The principle
goal has been to present the raw statistics necessary for tackling the
issue of thin disk survival in $\Lambda$CDM and for providing a
cosmological context for more focused simulations aimed at
understanding the role of mergers for processes like morphological
transformation, star formation triggering, and AGN fueling.

Our main results may be summarized as follows:

\begin{enumerate}

\item  Mass accretion into halos of mass $M_0$ at $z=0$ is dominated by
mergers with objects of mass
 $m  \simeq (0.03 - 0.3) M_0$ (Figure \ref{dfdm}).
Typically, $\sim 1-4$ mergers of this  size occur over a halo's history
(Figure \ref{NgtM}).   Because these mergers  tend  to occur when  the
main  progenitor's mass, $M_z$,  was somewhat smaller  than $M_0$, these
dominant events have fairly large {\em merger ratios},  $m/M_z
\simeq 0.1-0.6$ (see Appendix).

\item The mass accretion function, $n(m)$, 
of mergers larger than $m$ accreted over a halo's history 
 is well-described, on average, by
a simple analytic form, $n(x \equiv m/M_0) = A \, x^{-\alpha} \, (0.5-x)^{\beta}$, 
with $\alpha = 0.61$ and $\beta = 2.3$.
The normalization increases as a function of the halo's mass at $z=0$, $M_0$, as
$A \simeq 0.47 \log_{10}(M_0)-3.2$.  By extrapolating this fit, we find that
the total mass fraction accreted in objects of any mass ($m>0$)
does not asymptote to 1.0, but rather increases with
$M_0$ from $\sim 50 \%$ in $M_0 = 10^{11.5} \Msun$ halos to
$\sim 70 \%$ in $M_0 = 10^{13} \Msun$ halos.  This suggests that
a non-zero fraction of a halos mass may be accreted as truly ``diffuse''
material.

\item An overwhelming majority ($95 \%$) of Milky Way-sized halos 
with $M_0 \simeq  10^{12} \Msun$ have accreted  an object  larger than
the Milky Way's disk ($m \gtrsim 5 \times  10^{10} \Msun$) in the last
10 Gyr.  Approximately  $70 \%$ have had  accretions with $m > 10^{11}
\Msun$ objects over the same  period, and  $40 \%$  have had $m > 2
\times 10^{11} \Msun$ events (Figure \ref{MMfrac}).

\item Halo merger histories
 are approximately self-similar  in $m/M_0$ for  halos with  masses in  the range $M_0 =
 10^{11}-10^{13}  \Msun$ (Figure \ref{LMsincet}).   This suggests that
 the empirical trend  for late-type  galaxies to   be more common   in
 smaller halos is not governed by differences in merger histories, but
 rather is associated with baryon physics.

\item Typically, a small fraction, $\sim 20-30  \%$,
of a halo's final mass $M_0$
is accreted {\em after} the most recent large merger 
with $m > (0.1-0.2) \, M_0$ objects (Figure \ref{Macc}).  This suggests that
the ``regrowth'' of a disk from newly accreted material after a large merger 
 is unlikely.  Note that this does not rule out the possibility that a disk
reforms from gaseous material involved in the merger itself.

\end{enumerate}

The relatively high fraction of halos with large $m \sim 0.1 M_0$
merger events raises concerns about the survival of thin disk galaxies
within the current paradigm for galaxy formation in a $\Lambda$CDM
universe.  If we naively match percentages using Figure \ref{MMfrac},
we find that in order to achieve a $\sim 70 \%$ disk-dominated
fraction in $M_0 = 10^{12} \Msun$ halos, then $m \simeq 0.2 \, M_0
\simeq 2 \times 10^{11} \Msun \simeq 3 \times 10^{11} M_\odot$ objects
must not (always) destroy disks.  
Furthermore, since stars in the local thick disk and bulge 
are predominantly older than $\sim 10$ Gyr, this suggests that these
mergers in the past $\sim 10$ Gyr must not drive gas towards 
the bulge or significantly thicken the disk.
Note that the total mass in such an
accreted object is $\sim 10$ times that of the Milky Way disk itself.
Moreover halos typically do not accrete a significant fraction of
their final mass after these mergers ($\sim 20 \%$ on average).
Finally, as noted in the Appendix, $m \sim 0.2 \, M_0$ events
typically have merger {\em ratios} of $m/M_z \simeq 0.4$ at the time
of the merger.  These numbers do not seem encouraging for disk
survival, and may point to a serious problem with our 
current understanding of galaxy formation in a $\Lambda$CDM universe.

Our basic conclusion is unlikely to be sensitive to uncertain
cosmological parameters.  Note that the simulations considered here
have a fairly high $\sigma_8 = 0.9$.  At a fixed $\Omega_m$, a lower
$\sigma_8$ will systematically produce slightly {\em more} recent
merger events \citep[e.g.][]{zb03}.  However, given that the merger
fractions measured using $m/M_0$ are approximately self-similar in
$M_0$ (and therefore in $M_0/M_\star$), we expect that the overall
merger fractions will be fairly insensitive to power-spectrum
normalization.

As discussed in the introduction, a complete investigation into the
issue of disk survival will require an understanding of the orbital
evolution of objects once they have fallen within the main progenitor
halo's virial radius and on the subsequent impact of interacting
galaxies.  Both of these outcomes will depend sensitively on the
baryonic components in the main halo and in the smaller merging
object.  For this reason, the present work has focused on {\em
  halo mergers}, defined to occur when an infalling halo first crosses
within the virial radius of the main progenitor halo.  The merger
statistics presented here are relatively devoid of uncertainties and
can be used as a starting point for direct simulations of
galaxy-galaxy encounters.  Simulations of this kind will be essential
to fully address the broader implications of these frequent, large
mergers, which seem to pose a serious challenge to disk survival.

\acknowledgements The simulation used in this paper was run on the
Columbia machine at NASA Ames.  We would like to thank Anatoly Klypin
for running the simulation and making it available to us.  We are also
indebted to Brandon Allgood for providing the merger trees.  We thank
Chris Brook, T. J. Cox, Fabio Governato, Christian Maulbetsch, Brant
Robertson, Matias Steinmetz, Jerry Sellwood, and Rosemary Wyse for
helpful comments on an earlier draft, and Onsi Fakhouri and Chung-Pei
Ma for useful discussions about halo merger rates.  We thank the anonymous 
referee for several insightful suggestions  
that helped to improve the  quality and clarity of the paper.
JSB, RHW, and AHM
thank Rarija Mechslock for inspiration at an early phase of this
project.  JSB and KRS are supported by NSF grant AST 05-07916.  RHW
was supported in part by the U.S. Department of Energy under contract
number DE-AC02-76SF00515 and by a Terman Fellowship from Stanford
University.  ARZ is funded by the University of Pittsburgh. AHM
acknowledges partial support from a CUNY GRTI-ROUND 9 grant and from
an ROA supplement to NSF grant AST 05-07916.  \bibliography{stewart07}

\appendix

\begin{figure*}[b!]
  \begin{center}
   \subfigure{\includegraphics[width=.45\textwidth]{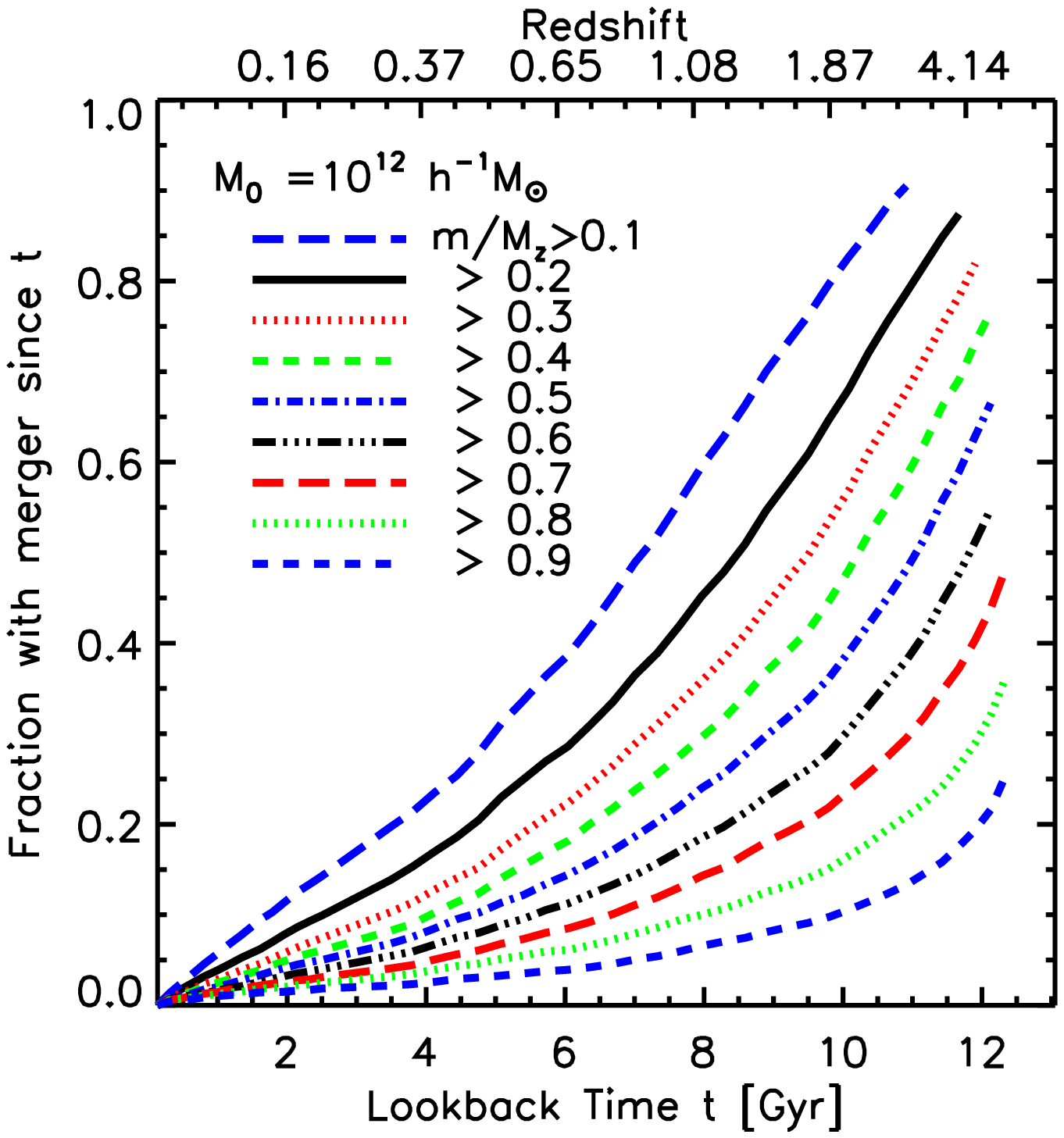}} 
    \subfigure{\includegraphics[width=.45\textwidth]{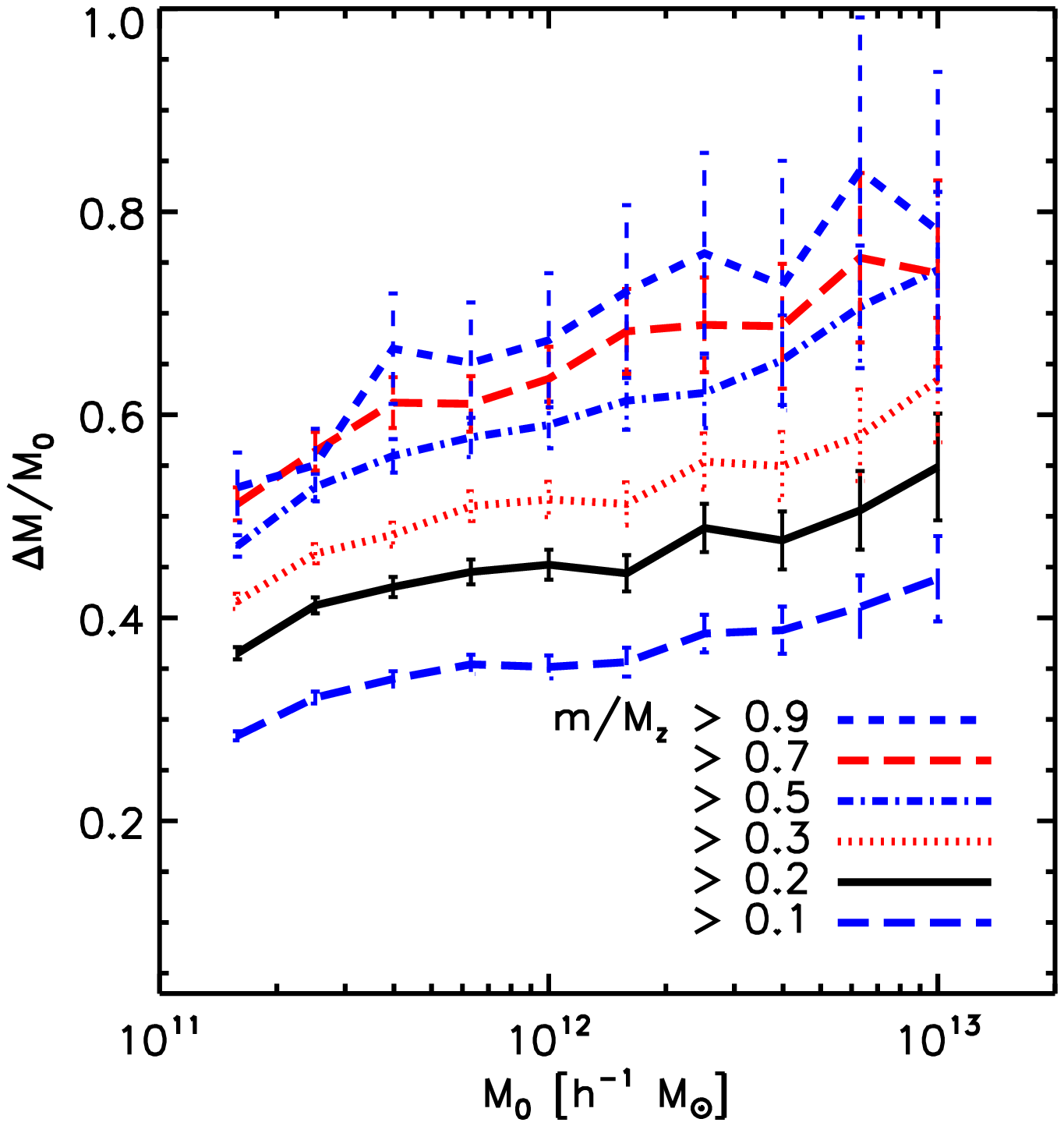}} 
    \caption{ 
{\em Left:} The fraction of Milky Way-sized halos, $M_0 \simeq 10^{12} \Msun$, 
that have  experienced  at  least  one  merger  larger  than a   given
{\em ratio}  $m/M_z$ since look-back time t, where $M_z$ is the
main progenitor mass {\em at the time of accretion}. Lines are truncated
at epochs where the mass resolution of the simulation limits our ability to resolve
a given $m/M_z$ ratio ({\em i.e.} when $M_z$ gets so small  that
a quoted $m/M_z$ ratio falls below the resolution with $m > 10^{10} \Msun$).
{\em Right:}
The    average change in mass $\Delta M/M_0$ that  a halo experiences
    since its last major merger.  Different lines show different
    ``major merger'' ratios, $m/M_z$, where the ratio is defined relative to
    the main progenitor mass {\em at the time of accretion}.  
   Error  bars  show Poissonian  $\sqrt{N}$
   errors  on the  number of host   halos averaged. 
}
\label{Afig1}
  \end{center}
\end{figure*}


The bulk of the paper focused on statistics of halo mergers at a fixed
absolute mass threshold $m$ on the merging objects.  In this section
we present statistics for $m/M_z$ --- the merger ratio relative to the
main progenitor mass $M_z$ at the redshift $z$ prior to accretion.   
By definition, $M_z \leq M_0$, causing merger statistics to show larger 
accretions when presented in terms of $m/M_z$.  For example, the merger ratio 
equivalent to Figure \ref{NgtM} (bottom panel) fits to $f(>x) = A x^{-\alpha} 
(1-x)^{\beta}$ instead of Equation \ref{eq:ngtm} (to better than $10\%$ across 
mass range $10^{11.5} - 10^{13.5} \Msun$), with parameters
$A(M_0) \simeq 0.05 \log_{10}(M_0)-2.0$, $\alpha = 0.04$,
and $\beta=2.0$, and $M_0$ in units of $\Msun$.  The $m/M_z$ equivalent for $n(>x)$ also
fits to this equation (to better than $10\%$ in the mass range $10^{12.0} -
 10^{13.5} \Msun$) with parameters $\alpha \simeq 0.4$, $\beta \simeq 2.0$, 
and $A(M_0) \simeq 2.0 \log_{10}(M_0)-21.5$.
Given that our simulation output has a fixed physical mass resolution,
$m_{\rm res} = 10^{10} \Msun$, and that halo main progenitor masses,
$M_z$, vary as a function redshift, the completeness in $m/M_z$ will,
in principle, vary from halo to halo as a function of each particular
mass accretion track and lookback time.  We have accounted for this
variable completeness limit in an average sense in what follows.

The {\em left} panel of Figure \ref{Afig1} shows the fraction of $M_0
= 10^{12} \Msun$ halos that have experienced {\em at least one} merger
larger than a given threshold ratio ($r_t = 0.1$, $0.2$, ... $0.9$) in
$m/M_z$ in the last $t$ Gyr.  This result may be compared to Figure
\ref{MMfrac}, which shows the analogous fraction computed using
fixed absolute $m$ cuts relative to $M_0$.  In order to provide
results that are robust to our completeness limit in $m$, for each
threshold cut, $r_t$, the lines are truncated at the lookback time
when the {\em average} $M_0 = 10^{12} \Msun$ halo's mass falls below
$M_z = m_{\rm res}/r_t$.  We see that $\sim 50 \%$ of halos have had a
$m/M_z > 0.4$ event in the last $\sim 10$ Gyr, and that $\sim 10 \%$
have experienced nearly equal-mass mergers, $m/M_z > 0.9$, in this
time.  These results are in good agreement with similar results quoted
by \cite{wyse04} for an analysis made by L. Hebb using the GIF
simulations.

The {\rm right} panel of Figure \ref{Afig1} shows that, typically, the
largest $m/M_z$ events occur {\em before} most of the final halo mass
$M_0$ is accreted.  Specifically we show the fraction of mass $\Delta
M/M_0$ accreted since the last major merger.  Results for different
major merger ratio thresholds are shown as different line types.  Each
line presents the {\em average} $\Delta M/M_0$ at fixed $M_0$ and we
only include halos that have actually had a merger of a given ratio
(within the last $\sim 11$ Gyr) in this figure.
Among halos that have had nearly
equal-mass merger events, $m/M_z \gtrsim 0.9$, the fraction of mass
that is accreted since that time is significant, with $\Delta M/M_0
\sim 70 \%$ for $M_0 = 10^{12} \Msun$ halos.  If we associate
post-merger accretion with the potential ``regrowth'' of a galactic
disk, then, by comparison with Figure \ref{Macc}, high merger-{\rm
  ratio} events are less of a concern for disk formation than high
$m/M_0$ ratio events.  Unlike most high $m/M_z$ events, mergers that
are large relative to $M_0$ typically have very little ($ \lesssim 20
\%$) fractional mass accretion after the merger.

\begin{figure}[tr!]
    \subfigure{\includegraphics[width=.45\textwidth]{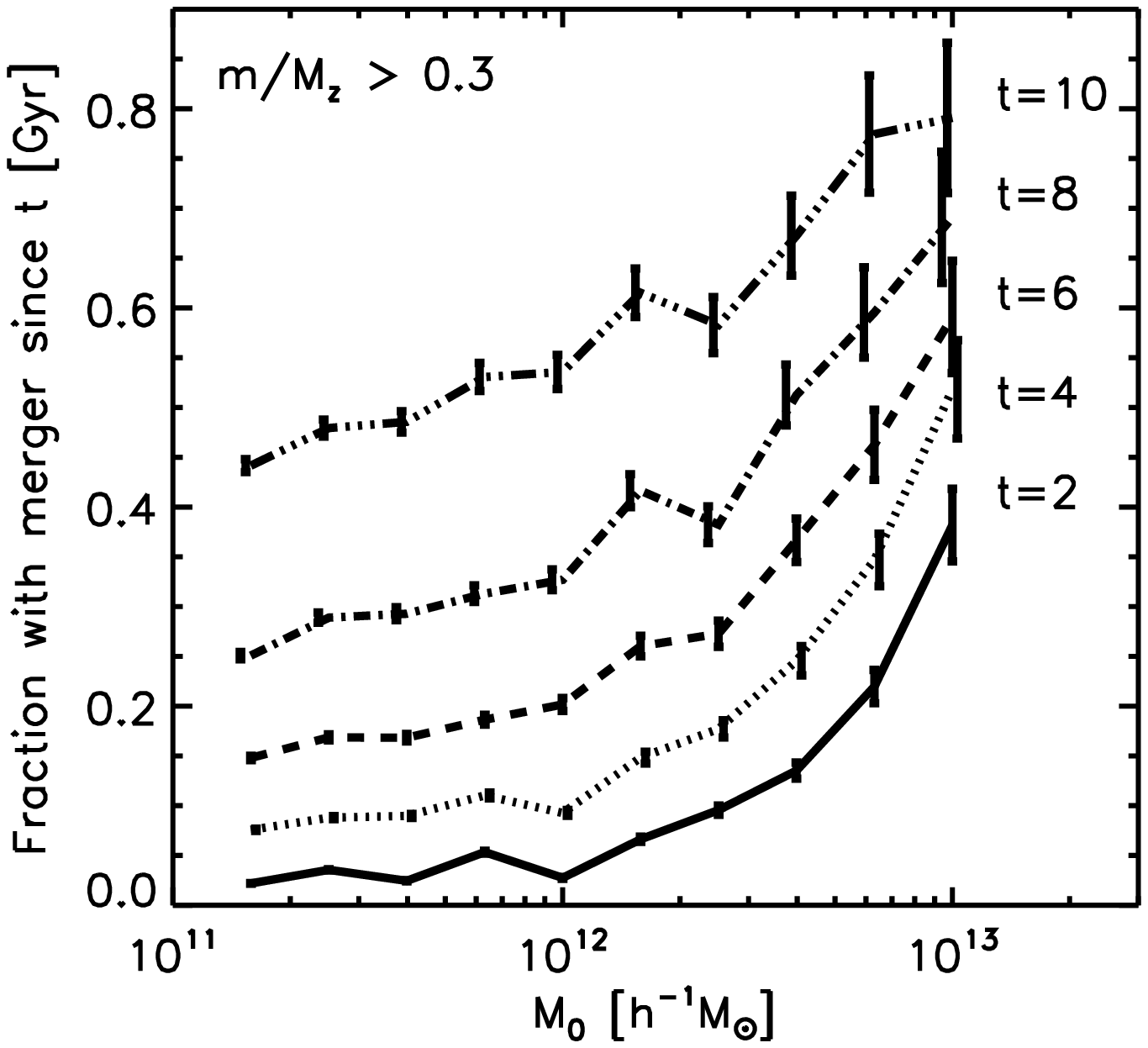}} 
\includegraphics[width=0.45\textwidth]{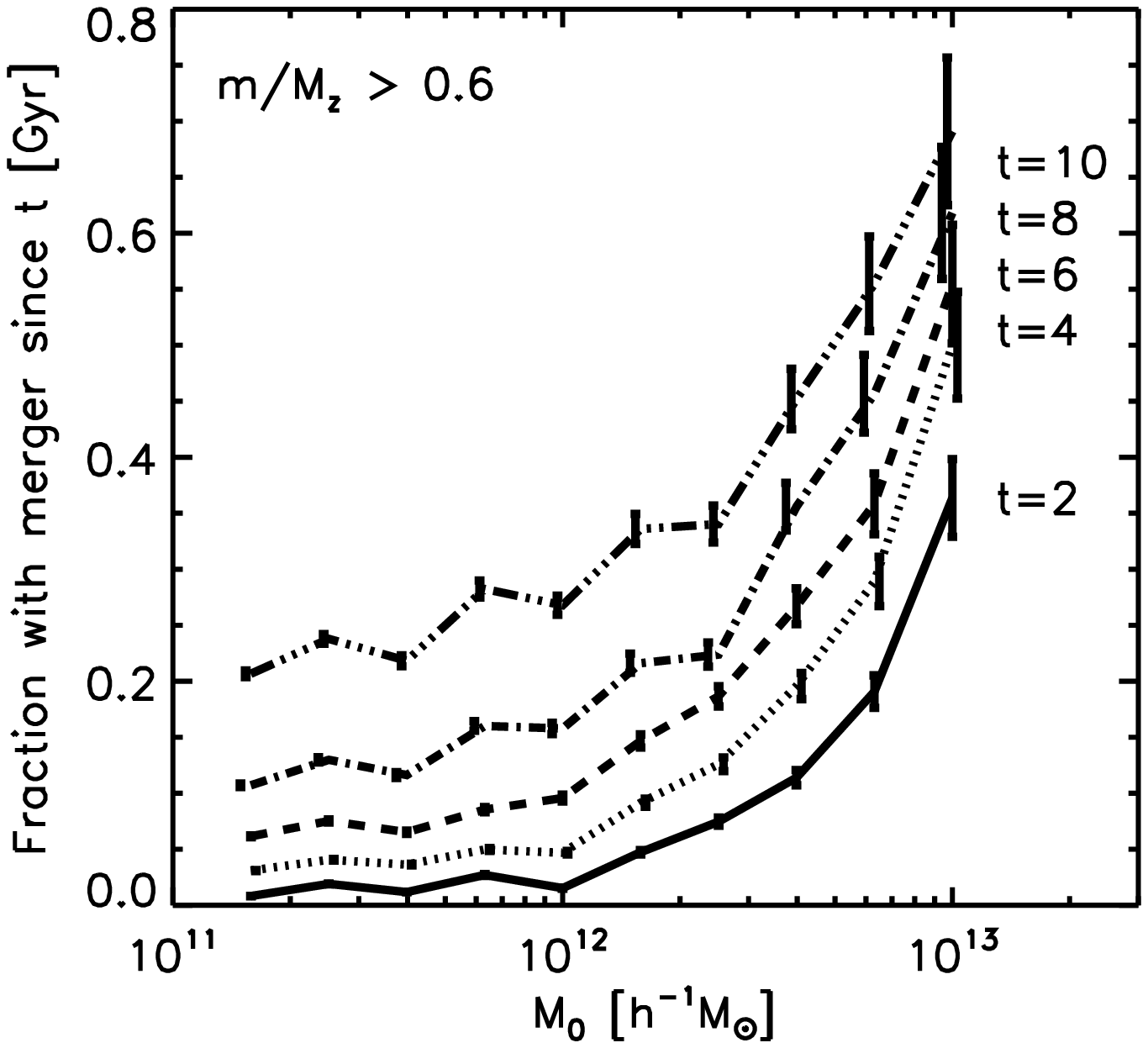} 
\caption{ 
{\em Left:}
The fraction of halos of mass $M_0$ at $z=0$ 
that have experienced a merger  ratio
larger than $m/M_z = 0.3$ in the last $t$ Gyr. 
The ratio is defined relative to the main progenitor mass at the
time prior to accretion $M_z$.
Error  bars are Poissonian based on the number of halos used in each
mass bin. Note the strong mass trend, in contrast to the results
presented in Figure \ref{LMsincet}, where mergers were defined on a
strict mass threshold, rather than merger ratio.
{\em Right:}
The fraction of halos of mass $M_0$ at $z=0$ 
that have experienced a merger  ratio
larger than $m/M_z = 0.6$ in the last $t$ Gyr. 
The ratio is defined relative to the main progenitor mass at the
time prior to accretion $M_z$.
Error  bars are Poissonian based on the number of halos used in each
mass bin. Note the strong mass trend, in contrast to the results
presented in Figure \ref{LMsincet}, where mergers were defined on a
strict mass threshold, rather than merger ratio.
}
\label{Afig2}
\end{figure}

The {\em left panel} of Figure \ref{Afig2} shows the fraction of halos
that have had a merger larger than $m/M_z = 0.3$ within the last $t =
2$, $4$, ... $10$ Gyr, as a function of halo mass $M_0$.  The {\em
  right panel} shows the same statistic computed for larger $m/M_z >
0.6$ mergers.  Unlike the result shown in Figure \ref{LMsincet}, there
is a fairly significant mass trend, with more massive halos more
likely to have experienced a major merger at a fixed lookback time.
Note, however, that most of these ``major mergers'' involve relatively
small objects in an absolute sense.  Much of this trend is driven by
the fact that $M_z$ falls off more rapidly with $z$ for high $M_0$
halos compared to low $M_0$ halos.

\begin{figure*}[t!]
  \begin{center}
\subfigure{\includegraphics[width=0.45\textwidth]{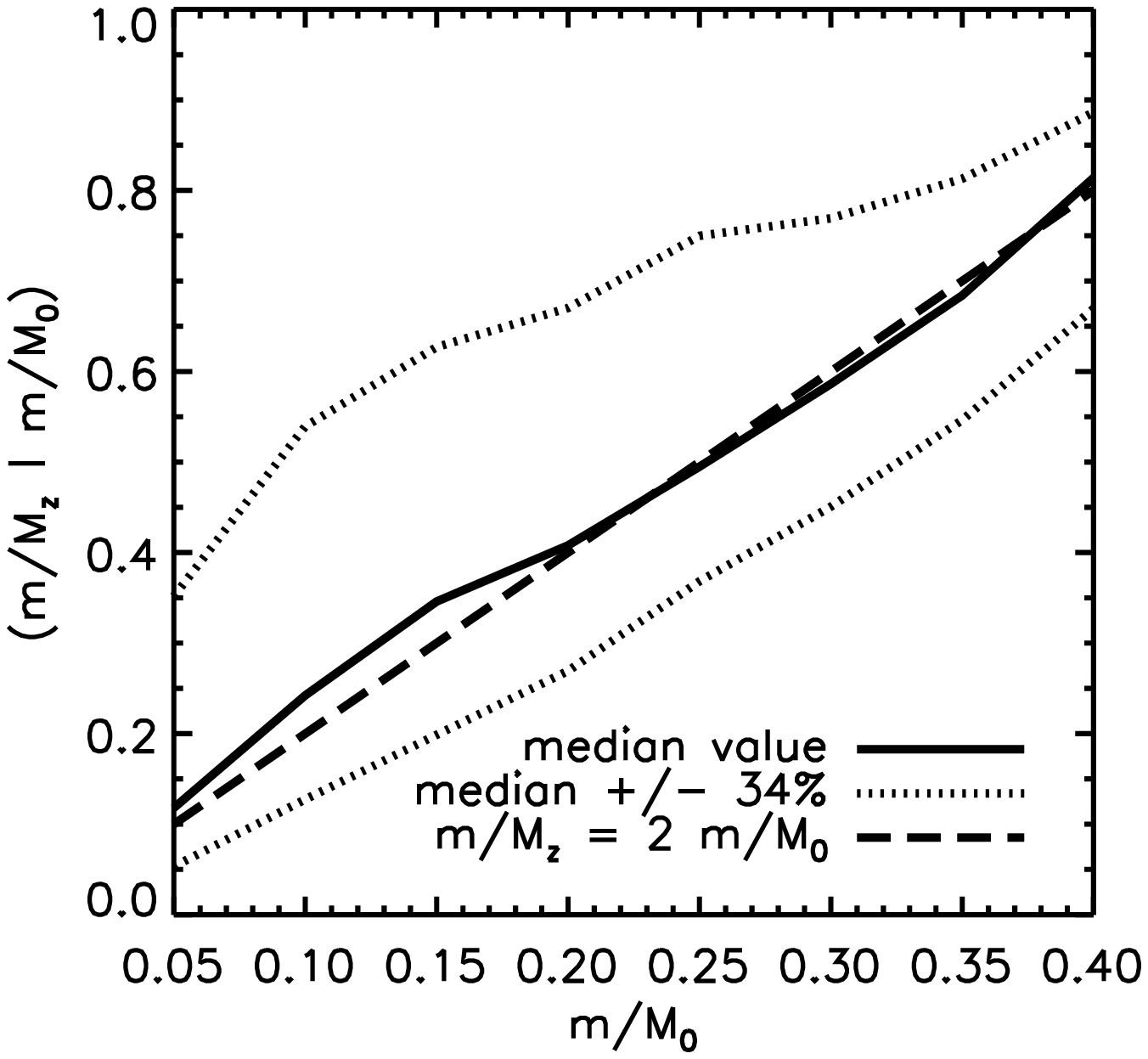}} 
\subfigure{\includegraphics[width=0.45\textwidth]{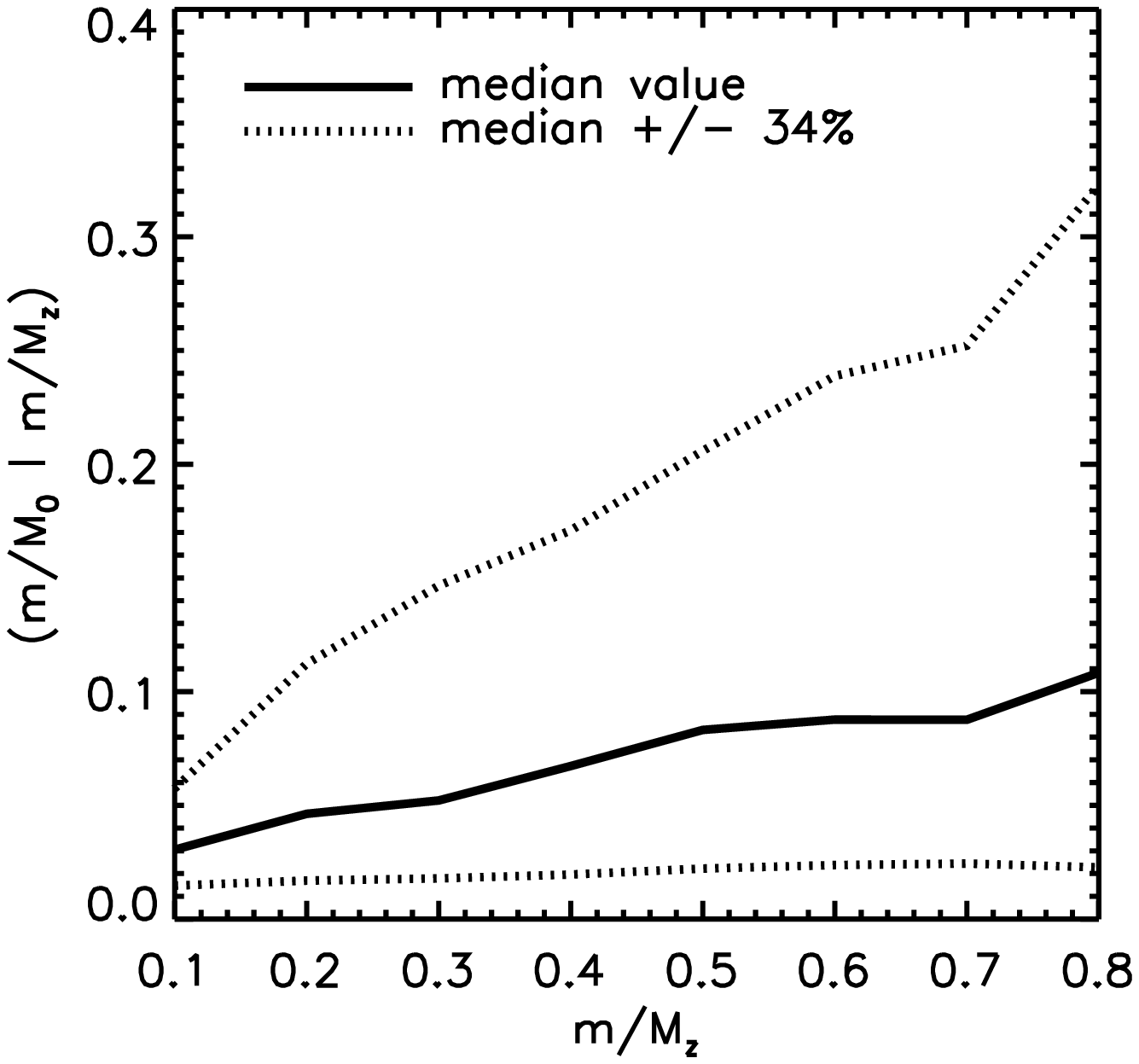}} 
\caption{ 
{\em Left:} 
An  illustration of $P(m/M_z|m,  M_0)$  --- the  distribution of merger
ratio  $m/M_z$ {\em at the  time of accretion}  given a value of
$m$ for $M_0 = 10^{12} \Msun$ halos.  We plot  $m/M_z$ vs. $m/M_0$ for
clarity.  The solid line shows  the median and  the dotted lines  show
the  $68 \%$ spread.   
{\em Right:} An illustration of $P(m/M_0|m/M_z)$ --- the distribution
of merging masses $m$ {\em given} a merger ratio $m/M_z$.  We see
that the majority of high $m/M_z$ events occur with $M_z$ is small
compared to the final halo mass $M_0$.  Therefore most high-mass ratio
mergers are small $m$ mergers in an absolute sense.}
  \label{Afig3}
\end{center}
 \end{figure*}

The previous discussions lead to explore the relationship between an
accreted halo's absolute mass $m$ and the mass-ratio it had when it
was accreted, $m/M_z$.  The two panels of Figure \ref{Afig3}
illustrate this relationship for objects accreted into $M_0 = 10^{12}
\Msun$ halos.  The thick, solid line in the {\em left} panel shows the
median merger ratio, $m/M_z$, that a halo of mass $m/M_0$ has when it
was accreted.  Specifically, we plot $m/M_z$ {\em given} $m/M_0$ in
this diagram.  We see that typically $m/M_z \simeq 2\, m/M_0$, as
shown by the dashed line in the figure.  The dotted lines show the 68
\% spread in the distribution of $m/M_z$ given $m/M_0$.  The opposite
relationship is shown in the {\em left} panel of Figure \ref{Afig3}.
Here, the thick, solid line shows the median $m/M_0$ value {\em given}
a merger of mass ratio $m/M_z$.  We see that the majority of
high-ratio events involve objects that are small compared to the final
halo mass $M_0$.  Note that this result, and the associated
distributions, are complete only to mergers that occur within the past
$\sim 11$ Gyr.  If we were able to track main progenitor masses $M_z$
back to arbitrarily early times, we would expect a very larger number
of high $m/M_z$ events with small absolute $m/M_0$ values.

\end{document}